# Infrared and Visual Interstellar Absorption Features toward Heavily-Reddened Field Stars [a]

M.G.Rawlings[1, 2], A.J.Adamson[3] and D.C.B.Whittet[4]

*(1) Centre for Astrophysics, University of Central Lancashire, Corporation Street, Preston, Lancs., PR1 2HE, U.K.*

*(2) Observatory, University of Helsinki, Kopernikuksentie 1, Tähtitorninmäki, FIN-00014, Finland.*

*(3) Joint Astronomy Centre. 660 N.A'Ohoku Place, Hilo Hawaii 96720 U.S.A.*

*(4) Department of Physics, Applied Physics & Astronomy, Rensselaer Polytechnic Institute, Troy, NY 12180, U.S.A.*

**ABSTRACT**

We present measurements of the 3.4-$\mu$m hydrocarbon dust absorption feature, and four visual diffuse interstellar bands, for twelve highly reddened ($9.0 < A(V) < 15.8$), early-type stars identified from the Stephenson (1992) catalogue, covering a range of Galactic longitude. The hydrocarbon feature is detected in eleven sightlines with optical depths of up to 0.05, and corresponding column densities are derived. The feature is weaker per unit $A(V)$ than previously reported, further underlining earlier observations of its enhancement in the Galactic Centre. A comparison of the 3.4-$\mu$m feature substructure indicates organic material similar to that seen in earlier diffuse medium studies, suggesting a uniform formation history across the Galactic disc. The profile is well matched by both meteoritic material and several laboratory-prepared analogues. The diffuse interstellar bands (DIBs) measured were $\lambda\lambda$6177, 6203, 6284 and 6614; no strong correlations were detected, either between these bands themselves or between the DIBs and the 3.4-$\mu$m feature, although $\lambda\lambda$6284 and 6614 appear to vary together. If the carriers of the DIBs are organic and molecular in nature and have cosmic carbon abundance requirements similar to those of the C-H stretch, these results imply that there is little direct carbon exchange between them and the aliphatic dust component. Measurements of the extinction to these objects suggest luminosity anomalies similar to that already seen in Cyg OB2 No. 12.

## 1  INTRODUCTION

The extinction and reddening effects of the diffuse interstellar medium (DISM) play a significant role in constraining the properties of modern dust models. Silicates and carbonaceous materials are the major components of many such models. The physical form and relative contributions of these components to the dust have recently been constrained by tightening of the cosmic abundance constraints (see Snow & Witt 1996 for a review), which appears to favour some forms of these materials over others (e.g. Mathis 1996; Li & Greenberg 1997). Of all the elements concerned, carbon imposes some of the strongest restrictions on current dust models.

Observations have suggested, some with greater confidence than others, the presence of a number of carbonaceous species in the ISM. Here we summarise these species and list what is known of their abundance:

<u>Solid-phase aliphatic hydrocarbons</u>. These produce an absorption feature at 3.4 $\mu$m identified with the DISM, being weak or absent in molecular clouds. Substructure in this feature at 3.38, 3.42 and 3.48 $\mu$m is identified with symmetric and asymmetric CH stretching modes. The relative strengths of these subfeatures implies a –CH$_2$–/–CH$_3$ ratio characteristic of short-chain hydrocarbons (Sandford et al. 1991 and Pendleton et al. 1994, Pendleton & Allamandola 2002). Details of the profile imply the presence of perturbing electronegative groups that can blend two of the four sub-features seen in normal alkanes. A good fit to the interstellar 3.4-$\mu$m profile was obtained by Greenberg et al. (1995), using solar-photolysed mixtures of simple laboratory-prepared ices. This, and the feature's apparent correlation with the 10-$\mu$m silicate feature (e.g. Sandford et al. 1995), was interpreted as evidence that the 3.4-$\mu$m carrier is photoprocessed organic mantles residing on silicate cores. However, the feature is unpolarized (Adamson et al. 1999), suggesting instead that small, unaligned grains, rather than organic mantles on silicate cores, may produce the feature.

---

[a] Based on data obtained with the UK Infrared Telescope, Mauna Kea Observatory, Hawaii, and the Isaac Newton Telescope, La Palma.

Tielens et al. (1996) estimate ~8 per cent of the cosmic carbon abundance to be locked up in hydrocarbon solids of some form, of which some or all is HAC (e.g. Jones, Duley & Williams, 1990). An indeterminate fraction of this material is probably in aliphatic bridges between aromatic structures (Pendleton & Allamandola 2002), and therefore capable of producing the 3.4-$\mu$m feature.

Gas-phase aromatic hydrocarbons (e.g. ionised or hydrogenated PAHs). The so-called Unidentified Infrared Bands (UIBs, UIR bands, also termed Infrared Emission Features, IEFs) ubiquitously seen in environments ranging from bright reflection nebulae through to Galactic cirrus clouds (e.g. Li & Draine 2002, Lemke et al. 1998, Boulanger 1998 & references therein) are now almost universally identified with gas-phase aromatic hydrocarbons. Inactive at optical wavelengths in their neutral form, the cations of classical PAHs also exhibit rich spectra across the wavelength ranges associated with the DIBs (Allamandola, Tielens & Barker, 1989). Probably the strongest evidence for this association is the wavelength correspondence between diffuse bands and emission bands detected in the Red Rectangle nebula (Fossey 1991, Sarre 1991 and Scarrott et al. 1992; Duley 1998). Some authors (e.g. Cami et al 1997, Moutou et al. 1999) have proposed that each DIB is produced by a different carrier species, with the so-called DIB families arising from similarities between some molecules. However, it is also quite easy to link a given DIB with a large range of possible PAH cations, and the relationship between DIBs and PAH emission in hot circumstellar regions is not conclusively in favour of a common set of carriers (e.g. Le Bértre & Lequeux 1993; Seab 1995). Variations in ionisation/hydrogenation states are a possible explanation. Salama et al. (1999) compared laboratory matrix isolation spectra of several PAH cations with the spectra of selected reddened early-type stars. They also found that although PAH ions are good carrier candidates for some of the DIBs, unambiguous identification is difficult, due to the effects of matrix shifts and feature broadening in the laboratory spectra. In an attempt to alleviate the matrix broadening/phase-shift problem, Krełowski et al. (2001) performed a comparison of gas phase spectra of the naphthalene cation with both high-resolution Echelle and conventional low-resolution astronomical spectra. Although reasonably good profile and width matching was obtained for two or three DIBs, an unexplained wavelength shift was detected. Future gas-phase spectroscopy may clarify the matter.

PAHs capable of producing the UIR bands are thought to contribute between 10 and 20 per cent (Allamandola et al. 1999, Dwek et al. 1997) of the available cosmic carbon.

Gas-phase aliphatic hydrocarbon chains. These species form the majority of polyatomic molecules identified in the gas phase in molecular clouds and circumstellar shells. Such molecules have been proposed as carriers of various radio emission lines (McCarthy et al. 2000) and at least some of the optical diffuse bands (Tulej et al. 1998). However, subsequent detailed spectral comparison work by Galazutdinov et al. 1999 and McCall et al. (2001) have cast serious doubt on this hypothesis, finding that features produced by the proposed species (e.g. $C_7^-$) poorly match measured astronomical profiles, intensity ratios and wavelengths. Motylewski et al. (2000) also compared gas-phase laboratory spectra of several gas-phase chain molecules with astronomical spectra and concluded that the astrophysical column densities are generally too low to produce the observed optical DIBs, although they suggested that some hydrocarbon chains could be responsible for several interstellar radio absorption features.

Carbon chains are expected to be less abundant than PAHs by a factor of 1000 (Allamandola et al. 1999).

Embedded aromatic molecules or their cations. Although frequently incorporated in dust models, either as a component of organic grain mantles in composite carbonaceous grains or as embedded cations in small HAC grains, observational evidence for such materials has been weak until recently. The predicted wavelengths of absorption features vary with molecular species (typically 3.25–3.29 $\mu$m). An aromatic C–H stretch feature 2–3 times weaker than the 3.4-$\mu$m feature is predicted near 3.28 $\mu$m (Pendleton et al. 1994). A possible detection of this feature toward the Galactic Centre was reported by Pendleton et al. (1994), and in the dense cloud regime, a tentative detection of a feature near 3.25 $\mu$m was reported by Sellgren et al. (1994) and subsequently seen by Chiar et al. 2000. More recently, Chiar et al. (2002) have produced detailed spectra of the Galactic Centre that confirm the presence of widespread aromatic absorption that is well correlated with the aliphatic features. In general, however, most models still lack the identification of any given DIB with a particular carrier molecule, but some laboratory work supports the idea that both aliphatic and aromatic molecules may be present in solid grains which, when shocked or thermally processed, are capable of liberating a wide range of carbon chains, aromatics and fullerenes into the gas phase (Scott & Duley 1996; Scott, Duley & Pinho 1997). Fractional abundance estimates of aromatic species embedded in HAC are difficult to uniquely quantify, as they depend strongly on the C/H ratio of the particular HAC under consideration (Tielens et al. 1996).

Fullerenes & related species. Since their discovery by Kroto et al. (1985), organic cage molecules, their isotopes and their hydrogenated or partially hydrogenated counterparts have been suggested as carriers for several diffuse ISM features, including the DIBs and the 217.5 nm extinction bump (Webster 1996 and references therein, McIntosh & Webster 1992; Webster 1995). However, there is no direct observational evidence for the presence of these molecules in interstellar space, although the $C_{60}^+$ ion is a possible carrier for the IR DIBs at 13175 Å and 11797 Å. $C_{60}^+$ was also associated with features at 9577 Å and 9632 Å (Foing & Ehrenfreund 1994), and although this identification was called into question by Jenniskens et al. (1997), it has received more recent support by Galazutdinov et al. (2000). Because fulleranes (hydrogenated fullerenes) are unstable even in solution, there are currently no laboratory measurements detailing the spectroscopic properties of these molecules. This also raises the question of their ability to survive in the ISM.

Approximately 25 per cent of the total carbon budget would need to be in fulleranes to account for the broad visual DIBs (Snow 1995).

Whichever species give rise to the DIBs, the relationship between free-floating molecules and hydrocarbon solids is one that can potentially be addressed by observation. It has variously been suggested that the UIB and DIB carriers may be liberated into the gas phase by interstellar shocks (Duley & Jones 1990, Duley & Williams 1988, Galazutdinov et al. 1998). O'Neill, Viti & Williams (2002) and Viti, Williams & O'Neill (2000) calculated that hydrogenated $C^+$ species arriving at diffuse cloud HAC grain surfaces could produce significant amounts of $CH_4$ and its derivatives. Since the abundance requirements of carbon in the solid and gas phases appear to be broadly equal, an observable anti-correlation of the DIBs with the solid-phase hydrocarbon bending and stretching features might be expected.

A measurement of this kind is observationally challenging, due to the difficulty of observing visual absorption bands in objects that are reddened sufficiently to produce measurable $3.4-\mu m$ absorption. This paper exploits a sample of heavily reddened lines of sight to carry out a study of this type.

## 2 THE STEPHENSON (1992) SURVEY

The objective-prism survey published by Stephenson (1992; hereafter S92) contained more than 400 sources claimed to exhibit extremely red colours but no photospheric TiO bands, suggestive of heavily reddened early-type stars. Such sources would be ideal for determination of the properties of the Galactic ISM at higher $A(V)$ than hitherto accessible. However, subsequent observations of comparatively small subsets of the catalogue (Creese et al. 1995, Imanishi et al. 1996, Goto et al. 1997) suggested that a majority of the sources were actually K- or M-type stars with only moderate reddening. The first large-scale spectroscopic study of the S92 catalogue (Rawlings, Adamson & Whittet 2000, hereafter paper I) included 6100–6900Å spectroscopy of 191 stars, and confirmed that 87 per cent of that subset are of spectral type K–M, and a further 3 per cent are S stars. Paper I revealed 15 sources to be of spectral type G or earlier, with extinctions between $A(V)$ of 6–16. Given the paucity of such high-extinction sightlines outwith the Galactic Centre (see Rawlings 1999 for a review), this new sample effectively doubles the number of objects available for extinction studies at high $A(V)$. This paper describes the first such studies.

Firstly, we quantify the strength of the 3.4-$\mu m$ aliphatic hydrocarbon feature at high $A(V)$, and look for variations in the feature's composition-driven substructure as a function of broad Galactic location. The early-type spectra already published in paper I were found to exhibit several of the stronger DIBs within their limited wavelength range. We therefore have a unique dataset that allows us to investigate both the infrared carbonaceous features and the optical DIBs in the same lines of sight.

## 3 OBSERVATIONS AND DATA REDUCTION

New infrared observations were carried out on the night of 1997 August 7 using CGS4 on the 3.8-m UK Infrared Telescope (UKIRT) at the Mauna Kea Observatory in Hawaii. To permit shorter readout and exposure times, the 256×32 pixel subarray was used. A north-south nod (along the slit) and '3×2' sampling was adopted. This configuration produced an effective wavelength range of 3.1–3.7$\mu m$. The weather was reasonably clear, but seeing was poor - averaging around 1 arcsecond, which results in spectral ripples generated by seeing fluctuations between samples and variable slit losses which reduce the photometric accuracy of the spectra.

Data were reduced using the Starlink packages CGS4DR, FIGARO and KAPPA. One-dimensional spectra extracted from the sky-subtracted frames were divided by synthetic ripple spectra to remove the seeing ripples. Cosmic-ray spikes in the spectra were interpolated across, and the spectra were linearly re-binned to identical wavelength ranges.

### 3.1 Atmospheric cancellation

Instrumental flexure causes sub-pixel spectral shifts over the course of long CGS4 exposures. Cross-correlation between the spectra and those of the atmospheric standards was used to determine the amount of shift necessary to correctly align the spectra before ratioing. Although the programme objects and standard stars were observed over a range of airmasses, it was found that airmass matching of programme stars with standards did not always produce the best cancellation of telluric bands, presumably reflecting rapidly-varying transmission. All of the programme star spectra were therefore cross-correlated and ratioed with all of the standard spectra to determine which combinations produced the best cancellation of telluric features. The best-matching standard spectrum *in all cases* was an observation of BS 7756 (HD 192985). This was rectified by ratioing with a polynomial continuum fit, and a synthetic template telluric absorption spectrum was generated, which could be linearly scaled to match the telluric bands for any given airmass; this is similar in spirit to the method introduced by Herbig (1975).

In general, it was found that the scaling factor required to optimally cancel the telluric bands at the long-wavelength end of the programme-star spectra (3.3–3.7$\mu$m) was slightly different from the optimal scaling factor for the short-wavelength end (3.1–3.3$\mu$m), where water lines dominate. For this reason, the two sections of the spectra were processed separately, with different scaling factors, and then merged to effect telluric cancellation for the whole wavelength range. Where appropriate, the resulting spectra were also smoothed.

BS 7756 is an F5V star; a blackbody effective temperature of 6530K was adopted when calibrating the ratioed spectra to produce the flux spectra shown in Figure 1. Figure 2 compares the spectrum of StRS 344 ratioed with its best airmass-matched standard, with the same spectrum ratioed with the rescaled, spliced standard spectrum to illustrate the improvement in telluric cancellation in the final spectra. At least six of these objects display hydrogen absorption in the Humphreys series, complicating the absorption profile. These features are indicated in Figure 3.

Optical depth spectra of the 3.4-$\mu$m feature were obtained (Figure 4) by ratioing the flux-calibrated spectra to polynomial continuum fits. The peak optical depth of the 3.4-$\mu$m hydrocarbon feature was then measured (Table 1); feature asymmetry was investigated using Gaussian fitting to the profiles (Section 5.1).

## 4   RESULTS AND ANALYSIS

The 3.4-$\mu$m feature was detected in eleven of the twelve sources, the only exception being the faint source StRS 432. The results are shown in Figure 5, which plots the correlation of the feature with $A(V)$ determined spectroscopically as in paper I. The new data suggest a systematically higher value of $A(V)/\tau(3.4\mu m)$ in regions relatively local to the Sun than determined by Sandford et al. (1995). By comparison with the Galactic Centre, this and the presence of at least $10^m$ of molecular cloud extinction in the latter line of sight (Lutz et al. 1996, Whittet et al. 1997) serves to further emphasize the known excess of hydrocarbon absorption in the Galactic Centre. To estimate column densities to these sources without sufficient signal-to-noise to determine the strengths of the individual subfeatures, we adopt a typical –$CH_2$– / –$CH_3$ ratio ~ 2.5 (Sandford et al. 1991, Pendleton et al. 1994) and weighted-average feature widths and absorption strengths of $\Delta v$ = 20.86 cm$^{-1}$ and A / (CH group) = 8.7 × 10$^{-18}$ cm group$^{-1}$ (using data presented by Sandford et al. 1991). The abundance relative to hydrogen was calculated using $N_H$ / $A(V)$ = 1.9 × 10$^{21}$ (Bohlin et al. 1978). These basic numerical results are included in Table 2.

As the 10-$\mu$m silicate feature correlates very similarly with $A(V)$ (Pendleton et al. 1994; Sandford et al. 1995), these results permit restrictions to be placed on which component(s) of the ISM may vary towards the Galactic Centre.

Since at least one of the Stephenson stars lies close to the line of sight to the Galactic Centre and exhibits more than 10$^m$ of visual extinction, Figure 5 implies that some of the extinguishing material is inefficient at producing the 3.4-$\mu$m feature. In practice it is apparently difficult to observe ~ 14$^m$ of extinction without encountering a significant amount of non-diffuse cloud material.

As shown in paper I, distances to these stars are difficult to determine, even in cases for which spectral types and luminosity classes are well constrained. Table 3 includes improved estimates of the distances to the StRS objects based on dereddened $R$-band fluxes from the visual spectra, including both upper and lower limits (columns 4 and 5 respectively) based on the limiting assumed spectral types and luminosity classes. These values supersede those of Table 3 in paper I, which are within the ranges quoted here. A comparison of these distances with estimates based on the simple assumption of an average extinction rate of 1.8 mag kpc$^{-1}$ (column 7) clearly illustrates the distance anomaly previously seen along other sightlines: the spectroscopically-derived distances are smaller in all cases. The relatively high visual extinction per unit distance inferred from the $R$-band fluxes might in some cases be reconciled with the measured extinctions if some fraction of the extinction towards the sources arises in clouds much denser than the diffuse ISM, or if these lines of sight sample a greater-than-average number of diffuse clouds per unit distance. The former holds for Galactic Centre sources, but the C–H stretch absorption features towards the StRS objects are very similar to those seen in the diffuse ISM.

## 5   STATISTICAL TESTS FOR VARIATIONS ON A GALACTIC SCALE

In Galactic distribution, the early-type Stephenson stars identified in paper I broadly divide into two groups, one towards the Galactic Centre (StRS 136 - 217, hereafter referred to as 'Low Longitude'), and the other in the approximate direction of Cygnus (StRS 344 - 432, 'High Longitude'). To test for systematic changes in the nature and absorbing efficiency of carbon grains as a function (to first order) of Galactocentric distance, the average feature strengths at 3.4 $\mu$m, were compared for these two groups. Statistical tests appropriate to comparisons of two small samples of two populations were applied to the measured feature strengths (normalised by total visual extinction). Upper limits, while numerically consistent with the rest of the data, were excluded from these tests. The remaining samples for each feature were tested for similar variances (using the $F$-test) and then similar means (using the $t$-test).

For $\tau$(3.4 $\mu$m)/$A(V)$, the null hypothesis that the two sample variances were the same could not be rejected at the 99 per cent confidence level. Both relevant forms of the $t$-test were therefore applied to the samples. For the two-tailed tests, the null hypothesis (that the two sample means were the same) could not be rejected at the 95 per cent confidence level. At this level, therefore, there is no justification for treating the samples at low and high longitude as two statistically separate populations.

### 5.1 Profile fitting and asymmetry of the 3.4-$\mu$m feature

Due to both the low signal-to-noise ratio and the presence in some cases of several Humphreys lines across the long-wavelength end of the hydrocarbon feature, a detailed decomposition of the CH vibrational sub-features is not possible. However, in almost all the lines of sight studied, the absorption appears to be dominated by a main feature at 3.4 $\mu$m with a wing feature at around 3.5$\mu$m, and we therefore parametrize the feature asymmetry by fitting two simple Gaussian absorptions.

The Starlink FIGARO routine GAUSS was used to fit two independent Gaussians. The results of the fits, characterized by the 'area' product of central depth and $\sigma$, are shown in Table 4 and Figure 6. Comparable fits to a number of laboratory and other astronomical hydrocarbon profiles are given for comparison (Figure 7). These include:

– EURECA A, B, C – Laboratory spectra of laboratory residues of photoprocessed low-temperature ices exposed to long-term solar UV radiation on the EURECA satellite (Greenberg et al. 1995)

– Core-mantle model – Calculated absorption by prolate (3:1) spheroidal silicate core-organic refractory mantle particles, using optical constants obtained from the EURECA data (Greenberg et al. 1995)

– GC IRS 6E, GC IRS 7 – the spectra of heavily-reddened Galactic Centre IR sources (Pendleton et al. 1994, Sandford et al. 1991)

– Murchison meteorite – Spectrum of organic residue acid-extracted from the primitive CM meteorite Murchison (Pendleton et al. 1994)

– Photolysed ice residue – Laboratory spectrum of a laboratory residue produced by the UV irradiation of a 10K $H_2$:$CH_3OH$:$NH_3$:CO = 10:5:1:1 interstellar ice analogue, followed by warm-up to 200 K (Pendleton et al. 1994, Allamandola et al. 1988)

– Ion bombarded methane – Spectrum of a laboratory residue produced by irradiating 10K methane ice with a 180 eV / C-atom dose of 75 keV protons (Pendleton et al. 1994)

– Quenched Carbonaceous Composite (QCC) – Laboratory spectrum of a room temperature filmy Quenched Carbonaceous Composite (QCC), synthesised from a hydrocarbon plasma (Pendleton et al. 1994, Sakata & Wada 1989)

– Seeded lab residue – Spectrum of a laboratory residue produced by the UV irradiation of a 10K $H_2O$:$CH_3OH$:$NH_3$:CO:$C_3H_8$ = 10:5:1:1:1 interstellar ice analogue, followed by warm-up to 200 K (Pendleton et al. 1994, Allamandola et al. 1988)

– Hydrogenated Amorphous Carbon (HAC) – Spectra of laboratory-synthesised HAC at 77 and 300K (Duley et al. 1998).

– CRL 618 – Astronomical spectrum of the fresh carbonaceous dust towards the protoplanetary nebula CRL 618 (Chiar et al. 1998)

– H-processed C nanoparticles – Spectra of laboratory-synthesised hydrogen-free HAC-like carbon nano-sized grains after exposure to atomic hydrogen (Mennella et al. 1999)

Some of the above samples, such as the EURECA samples and the meteoritic extracts, may be questionable as exact analogues of the diffuse ISM solid-phase carbonaceous carrier. These materials display a 5.83 $\mu$m CO stretch feature stronger than the 3.4-$\mu$m hydrocarbon stretch in laboratory spectra (Gibb & Whittet 2002), but this is not seen in the ISO spectrum of Cyg OB2 No. 12 (Whittet et al. 1997).

On visual inspection, the overall asymmetry of the 3.4-$\mu$m feature and the relative strength of the sub-features are not a strong function of environment; there is no evidence for variation with galactic longitude. There is a degree of uniformity in the feature asymmetry as measured by the ratio of the strengths of the two Gaussians (see Table 4). Excluding StRS 136 and 354 (due to poor signal-to-noise), and StRS 164 and 217 (due to significant hydrogen line contamination), the Stephenson stars exhibit asymmetry parameters greater than 2.0. This is dissimilar to the photolysed ice residue, but similar to most others, including ion-bombarded methane, QCC, warm HAC and H-processed C nanoparticles (Figure 8). Photolysed ices were also shown by Pendleton et al. (1994) to be a comparatively poor fit to their astronomical spectrum of GC IRS 6E. By implication, the chain lengths, abundance of electronegative perturbing groups and formation history of the hydrocarbon solids are fairly uniform across the Galaxy.

## 6 DIFFUSE INTERSTELLAR BANDS

As noted in Paper I, our visible spectra also contain several optical diffuse interstellar bands (DIBs), the majority of the proposed carriers of which incorporate organic molecules (see e.g. Fulara & Krełowski 2000, Krełowski & Greenberg 1999 for reviews of spectroscopic constraints on carriers).

Although several other DIBs appear to be present in one or two of the early-type spectral dataset, our primary objective here was to obtain as many measurements as possible for a commonly occurring set of bands. Consequently, only the features that were clearly distinguishable in all (or almost all) of the INT spectra were subjected to further study. This approach yielded four bands available for comparison with the 3.4-$\mu$m feature: $\lambda\lambda$6177, 6204, 6284 and 6614.

Band strengths were each measured three times for each source. The first of these measurements was conducted to be the most plausible in terms of continuum fitting and wavelength coverage. Upper and lower uncertainties were then estimated by re-measuring the bands with upper- and lower-limiting continua, and upper and lower plausible limits for the wavelength range of the band. In the very weakest cases, upper limiting values are quoted.

The strengths of the four selected DIBs in the early-type StRS spectra and Cyg OB2 No. 12 were measured using ABLINE. This program does not fit an assumed profile; instead, a numerical integration is performed based on the median wavelength, yielding band statistics, including central wavelength, depth, width, equivalent width (*W*) and asymmetry. The procedure first requires a user-generated polynomial continuum for spectrum normalization, which was here chosen to be of order 1-3 to minimize possible effects of spectra curvature. Manual limits may then be placed on the wavelength range of the spectral feature.

The resolution of the spectra does not permit a detailed study of the DIB profiles, and so consequently, full modelling of the band profiles to better eliminate blending effects was not attempted. However, additional care was taken in the measurement of the λ6177 band. This is a very broad, shallow band, sometimes exhibiting a red wing, and so significant blending with the narrow bands clustered around 6200 Å (e.g. λλ6196, 6203, 6205) is usually inevitable. Consequently, these features were interpolated across before any λ6177 measurements were made. The wavelength range of the band is also frequently difficult to determine accurately, and so plausible upper and lower estimates were again included in the error estimates. A strong He II line at 6170 Å has also been noted to cause blending problems in very hot stars, but this is not seen in our spectra - even those exhibiting a strong He I line at 6678 Å - and so is not considered further.

Blending is also an issue in the measurement of λλ6203 and 6284. The former sometimes suffers blending with another DIB centred at 6205 Å (Porceddu et al. 1991), and inspection of the Porceddu et al. (1991) spectra suggests that the two bands are approximately symmetric. λ6284 suffers from wing blending with the telluric molecular oxygen α band. To compensate for these effects, it was assumed that the relative contributions of each member of these blended pairs could be approximated by pairs of Gaussians (using the methods described above for the solid-phase hydrocarbon band) to estimate the relative contributions of the blended features. When measurably present, the fractional contributions from λ6205 and telluric oxygen were subtracted from the ABLINE values for λλ6203 and 6284 respectively. In almost all cases, these contributions were found to be within ~20%. Although the telluric oxygen band is known to exhibit a non-Gaussian profile, our interest was simply with estimating its fractional contribution to the blend, and this technique provides an approximation of this. The telluric oxygen and λ6284 features were fitted by Gaussians with average widths of 1.7 and 2.8 Å respectively. Any additional contribution to λ6203 from another DIB sometimes seen at 6196 Å (discussed in more detail in section 6.1) is a comparatively minor wing effect, and well within the reported uncertainties on the measurements.

Tables 5 and 6 list the measured strengths of both the 3.4-$\mu$m feature and the four DIBs present in the INT spectra (λλ6177, 6203, 6284 and 6614). Table 5 lists the measured band strengths, and Table 6 the band strengths per unit visual extinction. In both tables, italics denote upper limiting values only, and a blank entry in the τ(3.4 $\mu$m) column denotes a gap in the IR dataset. In the subsequent discussion, these latter quantities are the ones considered, as they are intrinsically more representative of the DIB properties.

As described above for the 3.4-$\mu$m measurements, the statistical *F*- and *t*-tests were applied to the two spatially separated 'high- and low-longitude' groups of objects for the normalised DIB strengths *W/A(V)*. Again, no statistical justification was found for treating the two groups as separate populations.

### 6.1 Relations between DIB strengths

Tests of DIB relationships in the context of these long, integrating lines of sight are shown in Figures 9 and 10[a]. No strong correlations were found. Firstly, there is no systematic dependence of either *W* or *W/A(V)* on *A(V)*. The strongest correlation is that between *W(6284)/A(V)* and *W(6614)/A(V)*, (with a correlation coefficient of 0.80). These two bands have not previously been identified as a family grouping.

This correlation does not support any particular familial relationships. λ6203 and λ6284 occupy family II of Chlewicki et al. 1986 and family II of Krełowski & Walker 1987. The expected clear correlation between these two bands is *not* seen in the StRS stars. To better test the proposed families, measurements should be made of corresponding members at shorter wavelengths. In particular, observations of λ5780 and λ5797 towards the Stephenson stars are needed to enable a discussion of cloud type independent of a given diffuse-ISM standard and family model.

More recently, Moutou et al. (1999) reported a strong correlation between λλ6614 and 6196, and this is clearly worthy of further investigation. Unfortunately, the number of features near 6200 Å, combined with the limited resolution of the spectra, precludes the establishment of the reliable continuum needed for the measurement for this band. Secondly, our primary objective was to provide a sample of DIBs that appeared in as many of the sample as possible, and the feature is very weak or not seen in at least eight of the spectra. Nevertheless, λ6196 does appear to be present towards at least some of our sightlines, and some of those also exhibit a relatively strong λ6614 band. However, inspection of the spectra in paper I suggests that λ6196 never appears to be more than half of the depth of λ6614, as suggested by the averaged schematic spectrum presented in Figure 8 of Weselak et al. (2001). Furthermore, the apparent lack of a λ6660 band in the Stephenson sightlines does not support the strong correlation between this and λ6614 reported by Weselak et al. (2001), lending weight to their assertion that the two bands arise from different (but possibly related) carriers. Of course, the Moutou et al. (1999) and Weselak et al. (2001) studies are based on sightlines that are not heavily reddened, but based on the Stephenson stars spectra obtained thus far, it appears that these correlations do not hold for higher $A(V)$.

---

[a] Measurements of the DIBs in the StRS objects were found to be similar to those of Cyg OB2 No. 12, and are shown together.

Some disagreements between proposed DIB groupings might be attributed to differences in the source-selection criteria of the various surveys. There are two approaches. The first, adopted by, e.g. Chlewicki et al. (1986) and this work, is to observe high-extinction lines of sight. This permits integration of the DIBs properties over a wide range of band strengths, path lengths and visual extinctions, but the sightlines' properties are often poorly known. The Stephenson stars inevitably lie behind a number of separate diffuse clouds whose physical conditions may vary significantly, and the Krełowski & Sneden (1995) $\zeta/\sigma$ cloud classification scheme is probably not appropriate. The second approach, adopted by, e.g. Krełowski & Walker (1987), Cami et al. (1997), Moutou et al. (1999) and Weselak et al. (2001) is to target single-cloud lines of sight whose properties are well understood. This only permits observations of DIBs for comparatively low extinctions, and. any DIB groupings identified may not hold for lines of sight with higher $A(V)$.

Bearing this in mind, we now address the mean nature of clouds along these lines of sight. Since the presence of dense clouds generally causes a weakening of the DIBs, and the StRS sightlines are long, a test of the diffuse/dense cloud contributions is important. We have limited wavelength coverage, so simply look for consistency with existing DIB family designations and the associated cloud types. For our limited dataset, this is determined solely by the relative strength of λ6614. DIBs in the Stephenson objects were compared with earlier measurements of DIB strengths in the 'purely' diffuse ISM, such as those summarised by Herbig (1995) and Jenniskens (2001)[a]. As the latter reference lists $W/E(B-V)$, a typical diffuse-ISM value of $R(V) = 3.05$ was used to convert to $W/A(V)$. Compared to the uncertainty in the DIB measurements the effect of a different choice of $R(V)$ is small. The results are listed in Table 7, and suggest that λλ6177, 6203 and 6284 have band strengths broadly consistent with primarily diffuse material. In contrast, λ6614, which is clear of telluric and photospheric features and not hard to measure, is of approximately half the expected strength towards the Stephenson stars (with an average value of $W/A(V) \sim 0.043 \pm 0.013$) when compared to the 'average diffuse medium' value of $0.076 \pm 0.012$ (Jenniskens, 2001). Ultimately, the significance of $W(6614)/A(V)$ depends on the adopted diffuse-ISM norm. If Jenniskens' relative DIB strengths for the diffuse ISM are correct, and one might prefer them to the Herbig (1995) values that rely on HD183143 alone, then the StRS sightlines (and that of HD183143) are consistent with the $\sigma$-type clouds of Krełowski & Sneden (1995). However, the Jenniskens (2001) strengths listed for λ6614 derive from one of the earliest DIBs surveys (Merrill 1934), the detailed accuracy of which may be questioned due to technical limitations of the time. Using the Herbig (1995) standards, the band ratios more closely resemble those in classical ($\zeta$-type) diffuse-ISM lines of sight.

## 7 COMPARISON OF THE DIBS WITH THE 3.4-$\mu$m FEATURE

There is a clear though indirect correlation between DIBs and the 3.4-$\mu$m feature: both appear to be weak or absent in dark clouds. However, correlation analyses of both the 3.4-$\mu$m feature and the DIBs in the same sources have not previously been possible, since most DIB studies have inevitably focused on lines of sight which exhibit comparatively low visual extinctions and undetectably weak 3.4-$\mu$m features. Since the Stephenson stars are bright both optically and in the NIR, and also heavily reddened, a combined investigation of both sets of features is now possible, albeit averaged over a long path length. Tables 5 and 6 show the equivalent widths of the four DIBs against the depth of the 3.4-$\mu$m C–H stretch feature. Figures 11 and 12 show correlations of measured feature strengths. As the StRS objects span a range of extinctions, the plots involving $W/A(V)$ (Figure 12) best reflect the average properties of the ISM along these lines of sight.

These data show no correlation between the C–H stretch and the DIBs. There is no relationship between DIB grouping and either the hydrocarbon feature strength or the visual extinction. Given that the strongest correlation seen in our data is between $W(6284)/A(V)$ and $W(6614)/A(V)$, we briefly consider the possibility that the hydrocarbon band strength might exhibit an accompanying trend. Figure 13 shows the apparent correlation between λλ6284 and 6614, and broadly indicates the corresponding measurements of the 3.4-$\mu$m feature. Taken together, these results suggest that whichever materials produce the measured DIBs do not directly exchange carbonaceous material with the solid-phase aliphatic hydrocarbons, and thus do not tend to support models which seek to produce the DIBs via organic molecules directly originating from the carbonaceous dust component. Conversely, this also argues against carbon-insertion processes for dust formation from gas-phase (DIB-producing) materials. To within errors, even cases with notably larger-than-average scatter from the $\tau – A(V)$ line show no obvious effect on the DIBs. Since similar cosmic carbon abundance requirements for both the DIBs carriers and the solid-phase aliphatics would result in an anti-correlation in the event of significant carbon exchange between the DIB carriers & the dust grains, our data imply one or both of the following conclusions, placing loose constraints on the DIB carriers:

1. The cosmic carbon requirements of the two species are in reality very different;

---

[a] http://www-space.arc.nasa.gov/~leonid/DIBcatalog.html

2. Material is not being exchanged between the two populations in large amounts.

While the lack of correlation between DIBs and 3.4$\mu$m absorption suggests a lack of exchange between gas and solid phases, it also implies that the fraction of extinction in dense and diffuse clouds along the sample of lines of sight studied here is relatively constant. However, there are hints: the tendency of Cyg OB2 to lie near the top of the diagrams in Figure 11 is consistent with the known lack of dense-cloud extinction in that line of sight. If there is a significant dense-cloud component towards other sightlines studied here, then a 3.0$\mu$m water ice feature might be detectable through further spectroscopy, and the associated dense-medium extinction measured. The long path lengths towards these objects could then be used to properly augment the comparison with the Galactic Centre.

## 8 CONCLUSIONS AND SUMMARY

Eleven early-type stars in the StRS catalogue have now been identified as exhibiting the 3.4-$\mu$m aliphatic C–H stretch absorption feature. These objects broadly separate into lines of sight toward the Galactic centre, and those at large longitude, towards Cygnus. Combining the infrared feature optical depths with the reddening estimates derived in Paper I has permitted an investigation of the hydrocarbon feature strength variation with $A(V)$. The results suggest a shallower slope of the $\tau - A(V)$ plot than previously observed (e.g. Sandford et al. 1995), but the 3.4-$\mu$m feature strength per unit extinction remains enhanced towards the Galactic Centre, particularly given the ~ $10^m$ non-diffuse cloud contribution to the ~ $31^m$ of extinction observed.

The hydrocarbon profile is broadly constant across the StRS sample, and is well represented by ion-bombarded methane, QCC, warm HAC and H-processed C nanoparticles. It is inconsistent with the photolysed ice residues. Detailed measurements of the feature asymmetry suggest minor variations in the relative strengths of the sub-features. These imply some secondary differences in the structure of the hydrocarbons, such as variations in carbon chain lengths, branching or attached electronegative perturbing groups. These variations do not depend on Galactic longitude.

For the first time, the strengths of several DIBs at visible wavelengths have been directly compared with that of the 3.4-$\mu$m feature. Four DIBs were studied: $\lambda\lambda$6177, 6203, 6284 and 6614. The bands do not statistically separate into two spatially separate groups. No strong mutual correlations were found between any of the DIBs studied, although $\lambda\lambda$6284 and 6614 do appear to vary together. The lack of strong detections of bands at $\lambda\lambda$6196 and 6660 indicates that their high correlations with $\lambda$6614 reported by Moutou et al. (1999) and Weselak et al. (2001) are not evident at high extinctions. Three ($\lambda\lambda$6177, 6203 and 6284) were found to be towards the low end of the ranges of strengths reported in the literature, but consistent within errors with the lowest values found in the literature. The strength per unit extinction of the remaining DIB ($\lambda$6614) relative to the diffuse ISM depends on the adopted norm. It was found to be weaker than the Jenniskens (2001) standard by a factor of typically ~ 2, but that standard itself differs from the Herbig (1995) standard by a similar factor. In the former case, the DIB strength ratios along the StRS sightlines are typical of the sigma clouds of Krełowski & Sneden, with an $R(V)$ ~ 4.5; in the latter they are more similar to those of classical diffuse-ISM lines of sight. These differences will be resolved by observations of the 5700 Å diffuse bands, the 3.0$\mu$m ice feature and near-IR photometry as a determinant of $R(V)$.

There were no detectable correlations or anti-correlations between the strengths of any of the four DIBs measured and that of the 3.4-$\mu$m C–H aliphatic stretch feature. This suggests either a lack of carbon exchange between the carriers of the DIBs and the 3.4-$\mu$m feature, and/or that the relative fractions of the cosmic carbon abundance required by the two populations are markedly different. This consequently places a loose constraint on current dust-gas interaction models.

## ACKNOWLEDGEMENTS

MGR acknowledges financial support from grants awarded by the UK PPARC and the Academy of Finland during the course of this research. AJA acknowledges financial support from the University of Central Lancashire, and by the UK PPARC. DCBW is supported by NASA grants NAG5-7598 and NAG5-7884. The INT is operated on the island of La Palma by the Isaac Newton Group in the Spanish Observatorio del Roque de los Muchachos of the Instituto de Astrofisica de Canarias. This research has made use of NASA' s Astrophysic Data System Bibliographic Services. The authors wish to thank the referee, Jaschek Krełowski, for his comments, which helped to improve the manuscript.

## REFERENCES

Adamson A. J., Whittet D. C. B. Chrysostomou A., Hough J. H., Aitken D. K., Wright G. S., Roche P. F., 1999, ApJ, 512, 224

Allamandola, L.J. Hudgins D.M., Bauschlicher Jr. C.W., Langhoff S.R., 1999, A&A 652, 659

Allamandola, L. J. Sandford, S. A., Valero G. J., 1988, Icarus, 76, 225

Allamandola L. J. Tielens A. G. G. M., Barker J. R. 1989, ApJS, 71, 733

Bohlin R. C., Savage B. D., & Drake J. F. 1978, ApJ, 224, 132

Boulanger F., 1998, in D'Hendecourt L., Joblin, C., Jones A. eds, Solid State Matter: The ISO Revolution, EDP, Springer, p. 19

Cami J., Sonnentrucker P., Ehrenfreund P., Foing B. H., 1997, A&A, 326, 822

Chiar J. E., Pendleton Y. J., Geballe T. R., Tielens A. G. G. M., 1998, ApJ, 507, 281

Chiar J. E., Tielens A. G. G. M., Whittet D. C. B., Schutte W. A., Boogert A. C. A., Lutz D., van Dishoeck E. F., Bernstein M. P., 2000, ApJ, 537, 749
Chiar J. E., Adamson A. J., Pendleton Y. E., Whittet D. C. B., Caldwell D. A., Gibb, E. L., 2002, ApJ, 570, 198
Chlewicki G., van der Zwet G. P., van Ijzendoorn L. J., Greenberg J. M., Alvarez P. P., 1986, ApJ, 305, 455
Creese M., Jones T. J., Kobulnicky H. A., 1995, AJ, 110, 268
Duley W. W., 1998, MNRAS, 301, 955
Duley W. W., Jones, A. P., 1990, ApJ, 351, L49
Duley W. W., Williams D. A., 1988, MNRAS, 230, 1P
Duley W. W., Scott A. D., Seahra S., Dadswell G., 1998, ApJ, 503, L183
Dwek E., et al., 1997, ApJ, 475, 565
Ehrenfreund P., Jenniskens P., 1995, in Tielens A. G. G. M., Snow T. P., eds, Proc. of The Diffuse Interstellar Bands, Kluwer, Dordrecht, p. 105
Fossey S. J., 1991, Nature, 353, 393
Fulara J., Krełowski J., 2000, NewAR, 44, 581
Galazutdinov G. A., Krełowski J., Musaev F. A., 1999, MNRAS, 310, 217
Galazutdinov G. A., Krełowski J., Musaev F. A., Ehrenfreund P., Foing B. H., 2000, 317, 750
Galazutdinov G. A., Krełowski J., Moutou C., Musaev F. A., 1998, MNRAS, 295, 437
Gibb E. L., Whittet, D. C. B., 2002, ApJ, 566, L113
Goto M., Sasaki Y., Imanishi M., Nagata T., Jones T. J., 1997, PASJ, 49, 485
Greenberg J. M., Li A., Mendoza-Gómez C. X., Schutte W. A., Gerakines P. A., de Groot M., 1995, ApJ, 455, L177
Herbig G. H., 1995, ARA&A, 33, 19
Herbig G. H., 1975, ApJ, 196, 129
Humphreys R. M., 1978, ApJS, 38, 309
Imanishi M., Sasaki Y., Goto M., Kobayashi N., Nagata T., Jones T. J., 1996, AJ, 112, 235
Jenniskens P., 2001, http://www-space.arc.nasa.gov/~leonid/DIBcatalog.html
Jenniskens P., Désert F. -X., 1994, ApJS, 160, 39
Jenniskens P., Mulas G., Porceddu I., Benvenuti P., 1997, 327, 337
Johnson H. L., 1966, ARA&A, 4, 193
Jones A. P., Duley W. W., Williams D. A., 1990, QJRAS, 31, 567
Krełowski J., Galazutdinov G. A., Musaev F.A, Nirski J., 2001, MNRAS, 328, 810
Krełowski J., Greenberg J. M., 1999, A&A, 346, 199
Krełowski J., Sneden C., 1995, in Tielens A. G. G. M., Snow T. P., eds, Proc. of The Diffuse Interstellar Bands, Kluwer, Dordrecht, p. 13
Krełowski J., Walker G. A. H., 1987, ApJ, 312, 860
Kroto H. W., Heath J. R., O'Brien S. C., Curl R. F., Smalley R. E., 1985, Nature, 318, 162
Le Bértre T., Lequeux J., 1993, A&A, 274, 909
Lemke D., Mattila K., Lehtinen K., Laureijs R. J., Liljeström T., Léger A., Herbstmeier U., 1998, A&A, 331, 742
Li A., Draine B. T., 2002, ApJ, 572, 232
Li A., Greenberg J. M., 1997, A&A, 323, 566
Lutz D., et al., 1996, A&A, 315, L269
McCall B. J., Thorburn J., Hobbs L. M., Oka T., York D. G., 2001, ApJ, 559, L49
McCarthy M. C., Chen W., Travers M. J., Thaddeus P., 2000, ApJS, 129, 611
Mathis J. S., 1996, ApJ, 472, 643
Mennella V., Brucato J. R., Colangeli L., Palumbo P., 1999, ApJ, 524, L71
Merrill P. W., 1934, PASP, 46, 206
Moutou C. Krełowski J., d'Hendecourt L., Jamroszczak J., 1999, A&A, 351, 680
Motylewski T., et al., 2000, ApJ, 531, 312
O'Neill P. T., Viti S., Williams D. A., 2002, A&A, 388, 346
Pendleton Y. J., Allamandola L. J., 2002, ApJS, 138, 75
Pendleton Y. J., Sandford S. A., Allamandola L. J., Tielens A. G. G. M., Sellgren, K., 1994, ApJ, 437, 683
Persi, P., Ferrari-Toniolo M., 1982, A&A, 111, L7
Porceddu I., Benvenuti P., Krełowski J., 1991, A&A, 248, 188
Rawlings M. G., 1999, PhD thesis, Univ. Central Lancashire
Rawlings M. G., Adamson A. J., Whittet D. C. B., 2000, ApJS, 131, 531, paper I
Sakata A., Wada S., 1989, in Allamandola L. J., Tielens A. G. G. M., eds, Proc. IAU Symp. 135, Interstellar Dust, Kluwer, Dordrecht, p. 191
Salama F., Galazutdinov G. A., Krełowski J., Allamandola L. J., Musaev F.A,, 1999, ApJ, 526, 265, 1999
Sandford S. A., Allamandola L. J., Tielens A. G. G. M., Sellgren K., Tapia M., Pendleton Y. J., 1991, ApJ, 371, 607
Sandford S. A., Pendleton Y. J., Allamandola L. J., 1995, ApJ, 440, 697
Sarre P. J., 1991, Nature, 351, 356
Scarrott S. M., Watkin S., Miles J. R., Sarre P. J., 1992, MNRAS, 255, 11P
Scott A., Duley W. W., 1996, ApJ, 472, L123
Scott A., Duley W. W., Pinho G. P., 1997, ApJ, 489, L193
Seab C. G., 1995, in Tielens A. G. G. M., Snow T. P., eds, Proc. of The Diffuse Interstellar Bands, Kluwer, Dordrecht, p. 129
Sellgren K., Smith R. G., Brooke T. Y., 1994, ApJ, 433, 179
Snow T. P., 1995, in Tielens A. G. G. M., Snow T. P., eds, Proc. of The Diffuse Interstellar Bands, Kluwer, Dordrecht, p. 379
Snow T. P., Witt A. N., 1996, ApJ, 468, L65
Sonnentrucker P., Cami J., Ehrenfreund P., Foing B. H., 1997, A&A, 327, 1215
Stephenson C. B., 1992, AJ, 103, 263, S92
Tielens A. G. G. M., Wooden D. H., Allamandola L. J., Bregman J. & Witteborn F. C., 1996, ApJ, 461, 210
Torres-Dodgen A. V., Tapia M., Carroll M., 1991, A&A Rev., 2, 167
Tulej M., Kirkwood D. A., Pachkov M., Maier J. P., 1998, ApJ, 506, L69
Viti S., Williams D. A., O' Neill P. T., 2000, A&A, 354, 1062
Wegner W., 1994, MNRAS, 270, 229
Weselak T., Fulara J., Schmidt M. R., Krełowski, J., 2001, A&A, 377, 677
Whittet D. C. B., 1992, Dust in the Galactic Environment, IOP Publishing, Bristol
Whittet D. C. B., et al., 1997, ApJ, 490, 729

TABLES

**Table 1.** Visual extinction and hydrocarbon feature optical depth towards the early-type Stephenson stars.

| StRS # | $A(V)$ [a] | $\tau(3.4\ \mu m)$ [b] |
|---|---|---|
| 136 | 14.6 ± 0.3 | 0.036 ± 0.007 |
| 158 | 11.5 ± 0.1 | 0.020 ± 0.004 |
| 164 | 13.0 ± 0.2 | 0.023 ± 0.008 |
| 173 | 7.2 ± 0.1 | 0.017 ± 0.009 |
| 185 | 6.6 ± 0.2 | 0.017 ± 0.009 |
| 217 | 12.7 ± 0.4 | 0.024 ± 0.007 |
| 344 | 13.1 ± 0.3 | 0.023 ± 0.005 |
| 354 | 13.7 ± 0.4 | 0.015 ± 0.008 |
| 371 | 15.8 ± 0.2 | 0.05 ± 0.02 |
| 375 | 9.2 ± 0.3 | 0.022 ± 0.008 |
| 392 | 10.1 ± 0.2 | 0.04 ± 0.01 |
| 432 | 9.0 ± 0.2 | 0.00 ± 0.04 [c] |

[a] Determined via INT spectroscopy; see paper I.

[b] Root mean square uncertainties on the continuum are represented in the errors on $\tau(3.4\ \mu m)$.

[c] $\tau(3.4\ \mu m)$ toward StRS 432 is subsequently assumed to have an upper limiting value of 0.04.

**Table 2.** Solid-phase hydrocarbon column densities towards the early-type Stephenson stars.

| StRS # | $N_{CH}$ / molecules cm$^{-2}$ | $N_{CH}$ / $A(V)$ | $N_{CH}$ / $N_H$ |
|---|---|---|---|
| 136 | $8.6 \times 10^{16}$ | $5.9 \times 10^{15}$ | $3.1 \times 10^{-6}$ |
| 158 | $4.8 \times 10^{16}$ | $4.2 \times 10^{15}$ | $2.2 \times 10^{-6}$ |
| 164 | $5.5 \times 10^{16}$ | $4.2 \times 10^{15}$ | $2.2 \times 10^{-6}$ |
| 173 | $4.1 \times 10^{16}$ | $5.7 \times 10^{15}$ | $3.0 \times 10^{-6}$ |
| 185 | $4.1 \times 10^{16}$ | $6.2 \times 10^{15}$ | $3.3 \times 10^{-6}$ |
| 217 | $5.8 \times 10^{16}$ | $4.6 \times 10^{15}$ | $2.4 \times 10^{-6}$ |
| 344 | $5.5 \times 10^{16}$ | $4.2 \times 10^{15}$ | $2.2 \times 10^{-6}$ |
| 354 | $3.6 \times 10^{16}$ | $2.6 \times 10^{15}$ | $1.4 \times 10^{-6}$ |
| 371 | $1.2 \times 10^{17}$ | $7.6 \times 10^{15}$ | $4.0 \times 10^{-6}$ |
| 375 | $5.3 \times 10^{16}$ | $5.8 \times 10^{15}$ | $3.1 \times 10^{-6}$ |
| 392 | $9.6 \times 10^{16}$ | $9.5 \times 10^{15}$ | $5.0 \times 10^{-6}$ |
| 432 (upper limit) | $9.6 \times 10^{16}$ | $1.1 \times 10^{16}$ | $5.8 \times 10^{-6}$ |

**Table 3.** Estimated distances to the early-type Stephenson stars.

| StRS # | Spectral Type [a] | Ref.[b] | d (Max.) [c] (kpc) | d (Min.) [c] (kpc) | $A(V)$ | d (scaled $A(V)$) [d] (kpc) |
|---|---|---|---|---|---|---|
| 136 | B8-A9I | J | 1.8 | 0.4 | 14.6 | 8.08 |
| 158 | O7-B2 | W | 1.5 | 0.2 | 11.5 | 6.39 |
| 164 | B8-A9I | J | 3.1 | 0.7 | 13.0 | 7.22 |
| 173 | O7-B2 | W | 3.5 | 0.5 | 7.2 | 4.00 |
| 174 | B5-B9.5 | W | 2.9 | 0.1 | 9.6 | 5.31 |
| 177 | Early? | ... | ... | ... | 14.9 | 8.28 |
| 185 | O6.5-B5 | W | 3.1 | 0.2 | 6.6 | 3.67 |
| 217 | B8-A9I | J | 2.8 | 0.6 | 12.7 | 7.03 |
| 344 | O8-B7 | W | 1.0 | 0.1 | 13.1 | 7.25 |
| 354 | O7-B3 | W | 1.0 | 0.1 | 13.7 | 7.58 |
| 368 | B6-B9.5 | W | 1.6 | 0.1 | 14.0 | 7.78 |
| 371 | Early | ... | ... | ... | 15.8 | 8.78 |
| 375 | O8-B9.5 | W | 3.5 | 0.1 | 9.2 | 5.11 |
| 392 | Early | ... | ... | ... | 10.1 | 5.61 |
| 432 | Early | ... | ... | ... | 9.0 | 5.00 |
| Cyg OB2 No. 12 [e] | B5Ia$^+$ | W | ... | 0.79 | 10.0 | 5.56 |

[a] See paper I for the spectral type determination method. Unless stated otherwise, luminosity classes are taken to be I-III.

[b] Intrinsic colours reference: 'J' denotes Johnson (1966), 'W' denotes Wegner (1994).

[c] Upper & lower limits on distance are derived from estimates of the *R*-band fluxes. All possible spectral type and luminosity class combinations suggested in column 3 were considered so as to obtain maximum and minimum values.

[d] Obtained by scaling $A(V)$ assuming an average Galactic disk extinction rate $< A(V) / L > = 1.8$ mag kpc$^{-1}$ (Whittet 1992).

[e] Assumed to have the intrinsic colours of a B5Ia star.

**Table 4.** Gaussian fitting parameters for the early-type Stephenson stars and various proposed interstellar organic analogues.

| Source / Sample | 3.4 μm Gaussian (Index 1) | | | | 3.5 μm Gaussian (Index 2) | | | | Asymmetry Parameter (G1/G2) |
|---|---|---|---|---|---|---|---|---|---|
| | λ (μm) | Height | σ | Area (G$_1$) | λ (μm) | Height | σ | Area (G$_2$) | |
| StRS 136 | 3.38 | 0.032 | 0.031 | 0.0025 | 3.46 | 0.021 | 0.037 | 0.002 | 1.3 |
| StRS 158 | 3.42 | 0.019 | 0.037 | 0.0017 | 3.49 | 0.009 | 0.018 | 0.0004 | 4.2 |
| StRS 164 | 3.40 | 0.019 | 0.028 | 0.0013 | 3.46 | 0.01 | 0.029 | 0.0008 | 1.7 |
| StRS 173 | 3.43 | 0.014 | 0.031 | 0.0011 | 3.49 | 0.005 | 0.021 | 0.0003 | 3.8 |
| StRS 185 | 3.42 | 0.016 | 0.034 | 0.0014 | 3.49 | 0.008 | 0.026 | 0.0005 | 2.6 |
| StRS 217 | 3.40 | 0.023 | 0.03 | 0.0018 | 3.47 | 0.019 | 0.029 | 0.0014 | 1.3 |
| StRS 344 | 3.40 | 0.026 | 0.031 | 0.002 | 3.48 | 0.015 | 0.025 | 0.0009 | 2.1 |
| StRS 354 | 3.41 | 0.014 | 0.023 | 0.0008 | 3.46 | 0.01 | 0.018 | 0.0005 | 1.9 |
| StRS 371 | 3.41 | 0.037 | 0.043 | 0.004 | 3.50 | 0.021 | 0.029 | 0.0016 | 2.5 |
| StRS 375 | 3.41 | 0.022 | 0.034 | 0.0019 | 3.48 | 0.01 | 0.016 | 0.0004 | 4.7 |
| StRS 392 | 3.39 | 0.035 | 0.037 | 0.0032 | 3.47 | 0.013 | 0.025 | 0.0008 | 4.0 |
| Core-mantle Model | 3.41 | 0.203 | 0.04 | 0.0205 | 3.50 | 0.095 | 0.024 | 0.0056 | 3.6 |
| EURECA A | 3.41 | 0.188 | 0.037 | 0.0176 | 3.50 | 0.098 | 0.029 | 0.0071 | 2.4 |
| EURECA B | 3.41 | 0.19 | 0.043 | 0.0203 | 3.50 | 0.099 | 0.024 | 0.006 | 3.4 |
| EURECA C | 3.41 | 0.171 | 0.038 | 0.0165 | 3.50 | 0.11 | 0.031 | 0.0086 | 1.9 |
| GC IRS 6E | 3.41 | 0.155 | 0.038 | 0.0146 | 3.49 | 0.067 | 0.02 | 0.0034 | 4.4 |
| GC IRS 7 | 3.41 | 0.161 | 0.032 | 0.0130 | 3.50 | 0.088 | 0.031 | 0.0069 | 1.9 |
| Murchison | 3.41 | 0.154 | 0.039 | 0.0152 | 3.50 | 0.074 | 0.019 | 0.0036 | 4.3 |
| Photolysed Ice Rsidue | 3.40 | 0.107 | 0.027 | 0.0071 | 3.47 | 0.133 | 0.049 | 0.0164 | 0.4 |
| Ion Bombarded Methane | 3.41 | 0.198 | 0.038 | 0.0187 | 3.49 | 0.104 | 0.026 | 0.0069 | 2.8 |
| QCC | 3.43 | 0.118 | 0.032 | 0.0095 | 3.51 | 0.06 | 0.024 | 0.0036 | 2.6 |
| Seeded Lab Residue | 3.42 | 0.147 | 0.039 | 0.0142 | 3.50 | 0.108 | 0.018 | 0.005 | 2.9 |
| HAC (300K) | 3.41 | 0.0292 | 0.039 | 0.0028 | 3.49 | 0.016 | 0.026 | 0.001 | 2.8 |
| HAC (77K) | 3.40 | 0.0171 | 0.031 | 0.0013 | 3.48 | 0.012 | 0.05 | 0.0014 | 0.9 |
| CRL 618 | 3.39 | 0.0852 | 0.031 | 0.0067 | 3.46 | 0.069 | 0.047 | 0.008 | 0.8 |
| H-processed C nanoparticles | 3.41 | 0.209 | 0.036 | 0.0187 | 3.49 | 0.098 | 0.027 | 0.0065 | 2.8 |

**Table 5.** A comparison of solid-phase hydrocarbon optical depth measurements and visual DIB equivalent widths. Upper limits are in italics.

| StRS # | A(V) | τ(3.4) | W(6177) | + | - | W(6203) | + | - | W(6284) | + | - | W(6614) | + | - |
|---|---|---|---|---|---|---|---|---|---|---|---|---|---|---|
| 136 | 14.6 ± 0.3 | 0.036 ± 0.007 | *0.68* | ... |  | 0.5 | 0.3 | 0.4 | 1.9 | 1.1 | 0.6 | 0.5 | 0.4 | 0.2 |
| 158 | 11.5 ± 0.1 | 0.020 ± 0.020 | 2.1 | 0.2 | 0.2 | 0.4 | 0.1 | 0.1 | 2.1 | 0.3 | 0.3 | 0.39 | 0.08 | 0.09 |
| 164 | 13.0 ± 0.2 | 0.023 ± 0.023 | 2.8 | 0.2 | 1.3 | 0.6 | 0.1 | 0.2 | 2.0 | 0.3 | 0.4 | 0.38 | 0.09 | 0.09 |
| 173 | 7.2 ± 0.1 | 0.017 ± 0.017 | 1.3 | 0.08 | 0.3 | 0.46 | 0.04 | 0.06 | 1.3 | 0.1 | 0.1 | 0.32 | 0.08 | 0.04 |
| 174 | 9.6 ± 0.2 | … | 1.46 | 0.46 | 0.18 | 0.4 | 0.1 | 0.1 | 1.8 | 0.2 | 0.2 | 0.3 | 0.1 | 0.1 |
| 177 | 14.9 ± 0.2 | … | *4.14* | ... |  | *2.06* | ... |  | 3.6 | 1.0 | 0.7 | 0.9 | 0.5 | 0.3 |
| 185 | 6.6 ± 0.2 | 0.017 ± 0.009 | 1.6 | 0.3 | 0.7 | 0.31 | 0.03 | 0.01 | 1.4 | 0.3 | 0.2 | 0.42 | 0.08 | 0.04 |
| 217 | 12.7 ± 0.4 | 0.024 ± 0.007 | 2.9 | 0.2 | 0.8 | 0.35 | 0.06 | 0.13 | 2.1 | 0.1 | 0.3 | 0.59 | 0.10 | 0.07 |
| 344 | 13.1 ± 0.3 | 0.023 ± 0.005 | 1.8 | 1.0 | 0.5 | 0.6 | 0.1 | 0.1 | 1.9 | 0.3 | 0.2 | 0.51 | 0.02 | 0.05 |
| 354 | 13.7 ± 0.4 | 0.015 ± 0.008 | *3.47* | 0.860 | 0.780 | 0.6 | 0.3 | 0.2 | 3.1 | 0.2 | 0.4 | 0.7 | 0.2 | 0.2 |
| 368 | 14.0 ± 0.1 | … | *5.70* | ... |  | 0.4 | 0.3 | 0.2 | 1.3 | 0.3 | 0.2 | 0.26 | 0.05 | 0.08 |
| 371 | 15.8 ± 0.2 | 0.05 ± 0.01 | *7.62* | ... |  | *0.71* | ... |  | 0.90 | 0.97 | 0.01 | 0.5 | 0.4 | 0.3 |
| 375 | 9.2 ± 0.3 | 0.022 ± 0.008 | 1.7 | 0.5 | 0.01 | 0.39 | 0.15 | 0.06 | 1.6 | 0.3 | 0.2 | 0.4 | 0.1 | 0.1 |
| 392 | 10.1 ± 0.2 | 0.04 ± 0.01 | 1.5 | 0.8 | 0.6 | 0.5 | 0.1 | 0.1 | 2.5 | 0.4 | 0.3 | 0.61 | 0.10 | 0.09 |
| 432 | 9.0 ± 0.2 | *0.04* | 2.4 | 0.2 | 1.2 | 0.24 | 0.07 | 0.07 | 1.5 | 0.2 | 0.4 | 0.35 | 0.07 | 0.04 |
| Cyg OB2 No. 12 | 10 ± 1 | 0.05 ± 0.01 | 2.4 | 0.2 | 0.4 | 0.62 | 0.06 | 0.06 | 2.3 | 0.2 | 0.2 | 0.37 | 0.09 | 0.02 |

**Table 6.** A comparison of solid-phase hydrocarbon optical depth measurements and visual DIB equivalent widths, all normalised with respect to *A(V)*. Upper limits are in italics.

| StRS # | *A(V)* | τ(3.4)/ *A(V)* | W(6177)/ A(V) | + | - | W(6203)/ A(V) | + | - | W(6284)/ A(V) | + | - | W(6614)/ A(V) | + | - |
|---|---|---|---|---|---|---|---|---|---|---|---|---|---|---|
| 136 | 14.6 ± 0.3 | 0.025 ± 0.0005 | 0.05 | … | | 0.03 | 0.02 | 0.03 | 0.13 | 0.08 | 0.04 | 0.04 | 0.03 | 0.01 |
| 158 | 11.5 ± 0.1 | 0.0017 ± 0.0003 | 0.19 | 0.02 | 0.02 | 0.04 | 0.01 | 0.01 | 0.19 | 0.02 | 0.02 | 0.034 | 0.007 | 0.008 |
| 164 | 13.0 ± 0.2 | 0.0018 ± 0.0006 | 0.21 | 0.01 | 0.10 | 0.04 | 0.01 | 0.01 | 0.16 | 0.02 | 0.03 | 0.029 | 0.007 | 0.007 |
| 173 | 7.2 ± 0.1 | 0.002 ± 0.001 | 0.20 | 0.06 | 0.06 | 0.06 | 0.01 | 0.01 | 0.18 | 0.02 | 0.02 | 0.04 | 0.01 | 0.01 |
| 174 | 9.6 ± 0.2 | … | 0.15 | 0.05 | 0.02 | 0.04 | 0.01 | 0.01 | 0.18 | 0.02 | 0.02 | 0.03 | 0.01 | 0.01 |
| 177 | 14.9 ± 0.2 | … | *0.28* | … | | *0.14* | … | | 0.24 | 0.07 | 0.05 | 0.06 | 0.03 | 0.02 |
| 185 | 6.6 ± 0.2 | 0.003 ± 0.001 | 0.24 | 0.04 | 0.11 | 0.047 | 0.005 | 0.002 | 0.21 | 0.03 | 0.02 | 0.06 | 0.01 | 0.01 |
| 217 | 12.7 ± 0.4 | 0.0019 ± 0.0006 | 0.23 | 0.017 | 0.07 | 0.028 | 0.005 | 0.010 | 0.16 | 0.01 | 0.02 | 0.047 | 0.008 | 0.006 |
| 344 | 13.1 ± 0.3 | 0.0018 ± 0.0004 | 0.13 | 0.08 | 0.03 | 0.04 | 0.01 | 0.01 | 0.15 | 0.02 | 0.01 | 0.039 | 0.002 | 0.004 |
| 354 | 13.7 ± 0.4 | 0.0011 ± 0.0006 | 0.25 | 0.06 | 0.06 | 0.05 | 0.02 | 0.02 | 0.23 | 0.02 | 0.03 | 0.05 | 0.02 | 0.01 |
| 368 | 14.0 ± 0.1 | … | *0.41* | … | | 0.03 | 0.02 | 0.02 | 0.10 | 0.02 | 0.02 | 0.019 | 0.004 | 0.006 |
| 371 | 15.8 ± 0.2 | 0.0031 ± 0.0006 | *0.48* | … | | *0.05* | … | | 0.057 | 0.062 | 0.001 | 0.03 | 0.03 | 0.02 |
| 375 | 9.2 ± 0.3 | 0.0023 ± 0.0009 | 0.19 | 0.05 | 0.01 | 0.043 | 0.016 | 0.006 | 0.17 | 0.03 | 0.03 | 0.05 | 0.01 | 0.01 |
| 392 | 10.1 ± 0.2 | 0.004 ± 0.001 | 0.15 | 0.08 | 0.06 | 0.05 | 0.01 | 0.01 | 0.24 | 0.04 | 0.03 | 0.06 | 0.01 | 0.001 |
| 432 | 9.0 ± 0.2 | *0.004* | 0.27 | 0.03 | 0.13 | 0.03 | 0.01 | 0.01 | 0.16 | 0.02 | 0.05 | 0.039 | 0.008 | 0.005 |
| Cyg OB2 No. 12 | 10 ± 1 | 0.006 ± 0.001 | 0.24 | 0.02 | 0.04 | 0.06 | 0.01 | 0.01 | 0.23 | 0.03 | 0.03 | 0.037 | 0.010 | 0.004 |

**Table 7.**[a] A comparison of literature measurements of A(V)-normalised DIB equivalent widths obtained from the literature and this work.

| Line of Sight | $W(6177)/A(V)$ | $W(6203)/A(V)$ | $W(6284)/A(V)$ | $W(6614)/A(V)$ | Reference |
|---|---|---|---|---|---|
| 'Average Diffuse' | 0.253 ± 0.046 | 0.035 ± 0.005 | 0.203 ± 0.016 | 0.076 ± 0.012 | Jenniskens (2001) |
| 'Diffuse' | 0.16 ± 0.07 | ... | 0.39 ± 0.16 | ... | Ehrenfreund and Jenniskens (1995) |
| HD 183143 [b] | 0.379 ± 0.052 | 0.044 ± 0.006 | 0.232 ± 0.022 | 0.107 ± 0.014 | Jenniskens (2001) |
| HD 183143 [b] | ... | ... | ... | 0.083 ± 0.001 | Sonnentrucker et al. (1997) |
| HD 183143 [b] | 0.305 ± 0.028 | ... | 0.357 ± 0.035 | ... | Ehrenfreund and Jenniskens (1995) |
| HD 183143 [b] | 0.612 | 0.083 | 0.498 | 0.092 | Herbig (1995) |
| HD 183143 [b] | 0.474 | 0.110 | 0.512 | 0.090 | Whittet (1992) |
| StRS Average | 0.20 ± 0.04 | 0.04 ± 0.01 | 0.17 ± 0.05 | 0.04 ± 0.01 | This work |

[a] $R(V)$ was assumed to be 3.05 in all cases.

[b] $E(B-V) = 1.24 \pm 0.08$ (Jenniskens & Désert, 1994) was assumed for HD 183143.

FIGURES

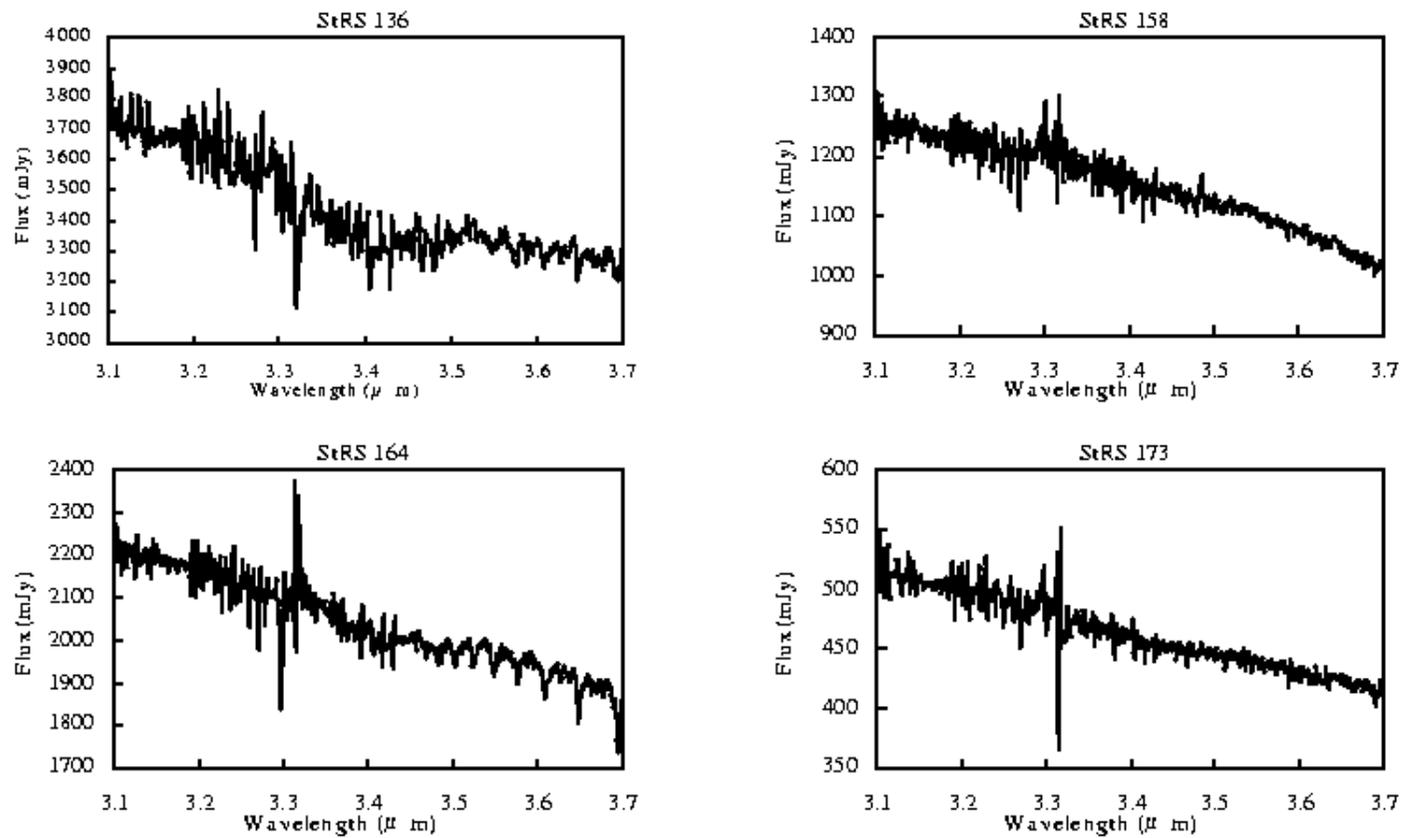

**Figure 1.** Calibrated 3.4 μm spectra of selected genuinely early-type Stephenson stars.

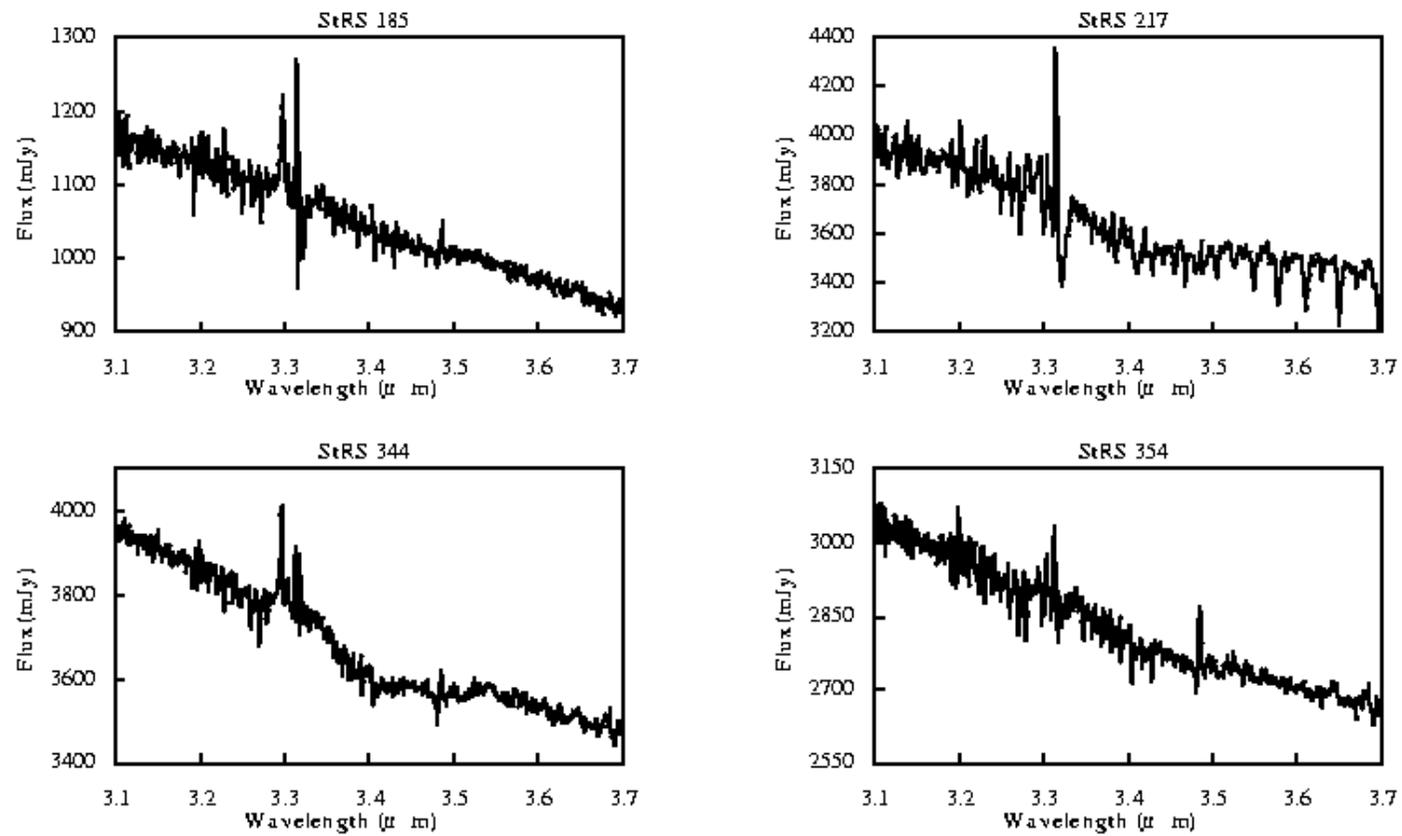

**Figure 1.** Continued

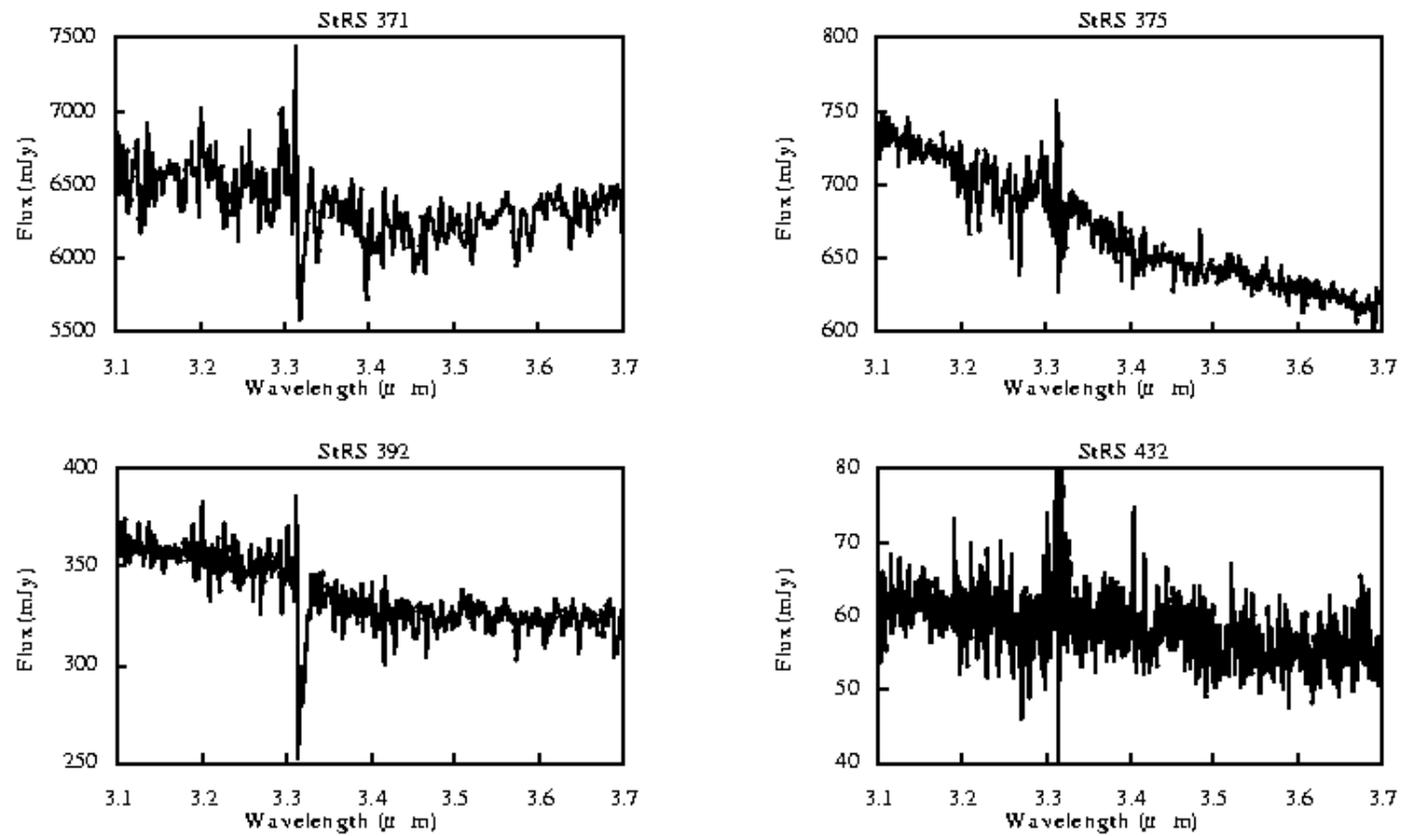

**Figure 1.** Continued

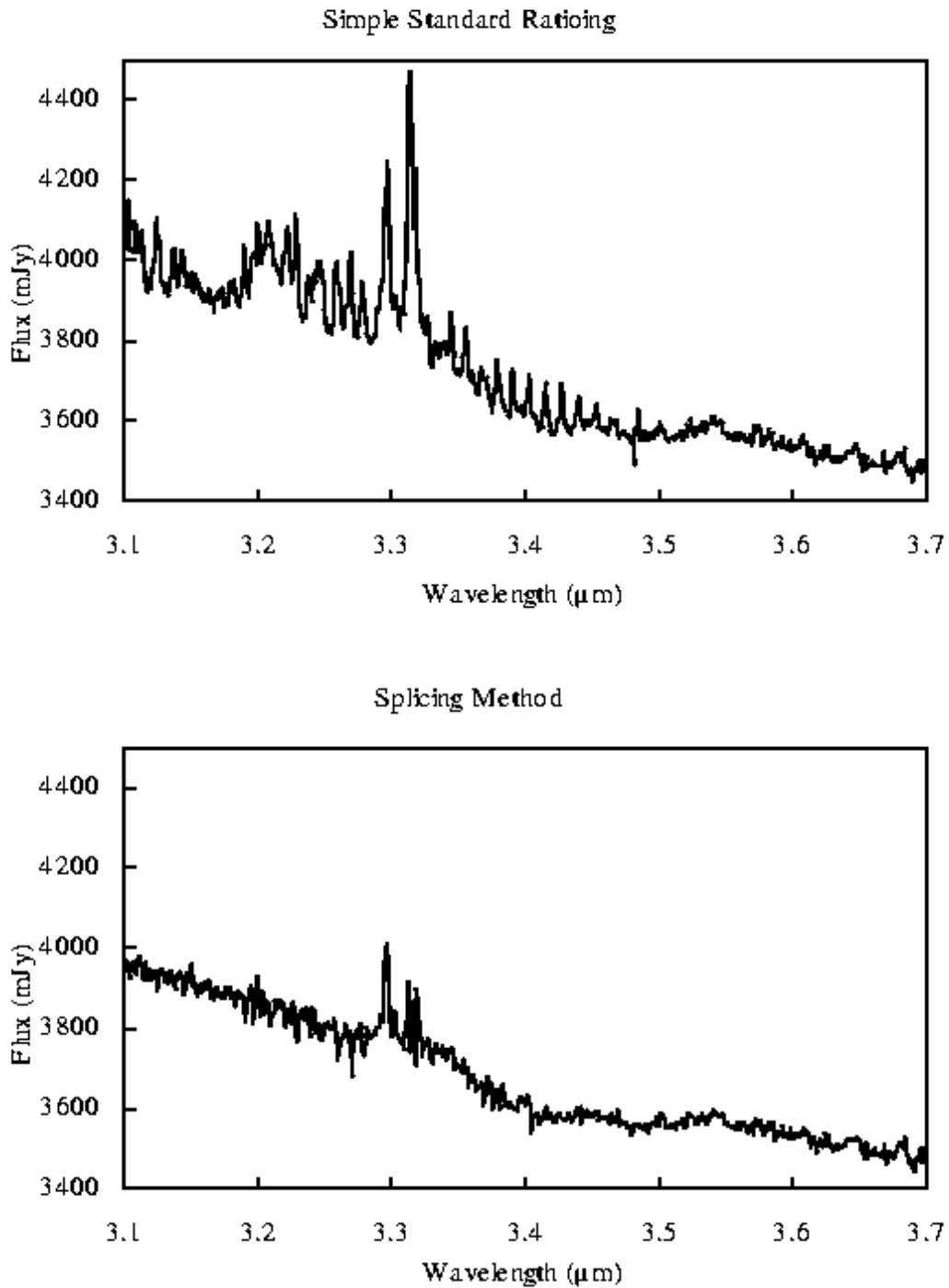

**Figure 2.** An example of the improvement in telluric band cancellation for the early-type StRS objects. The upper figure shows the spectrum of StRS divided by the best airmass-matched standard, and the lower figure shows the effect of using a rescaled, spliced standard.

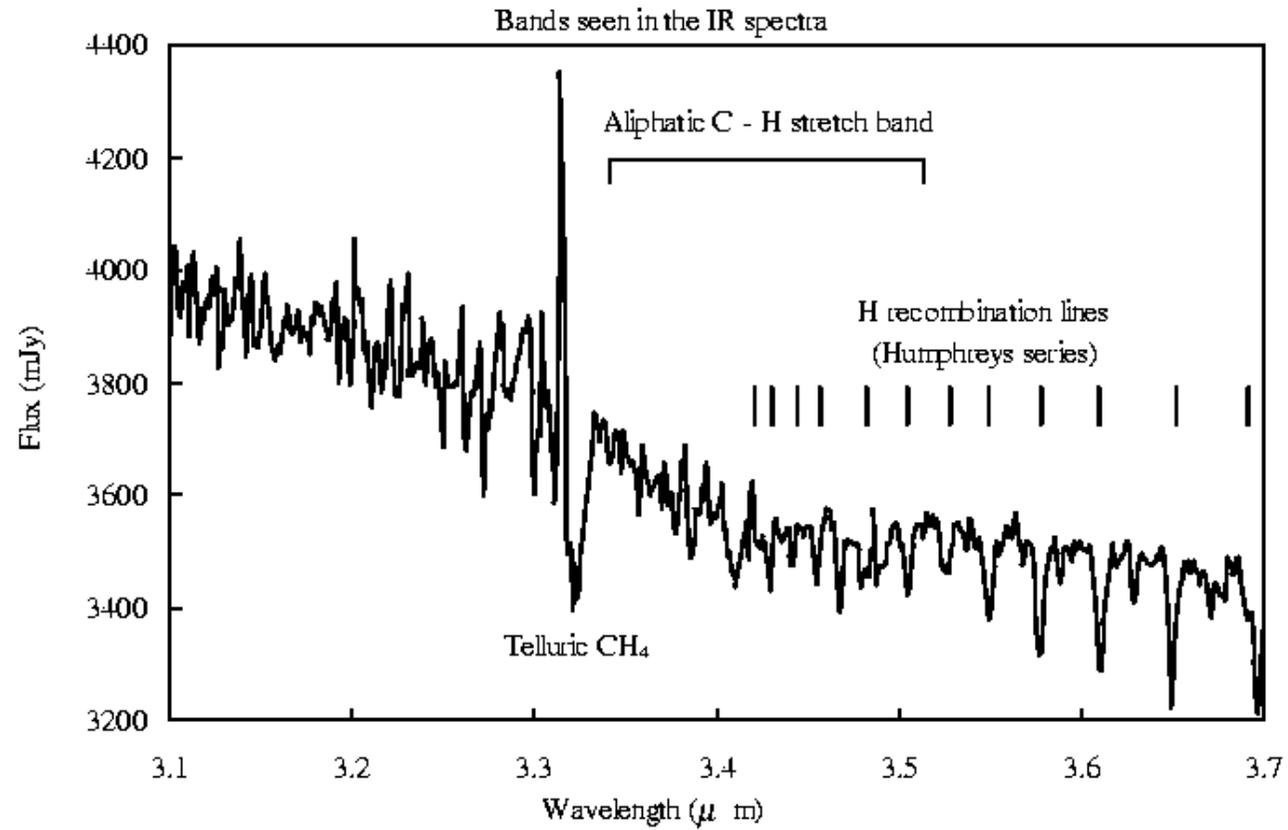

**Figure 3.** Bands detected in the IR spectra of selected early-type Stephenson stars.

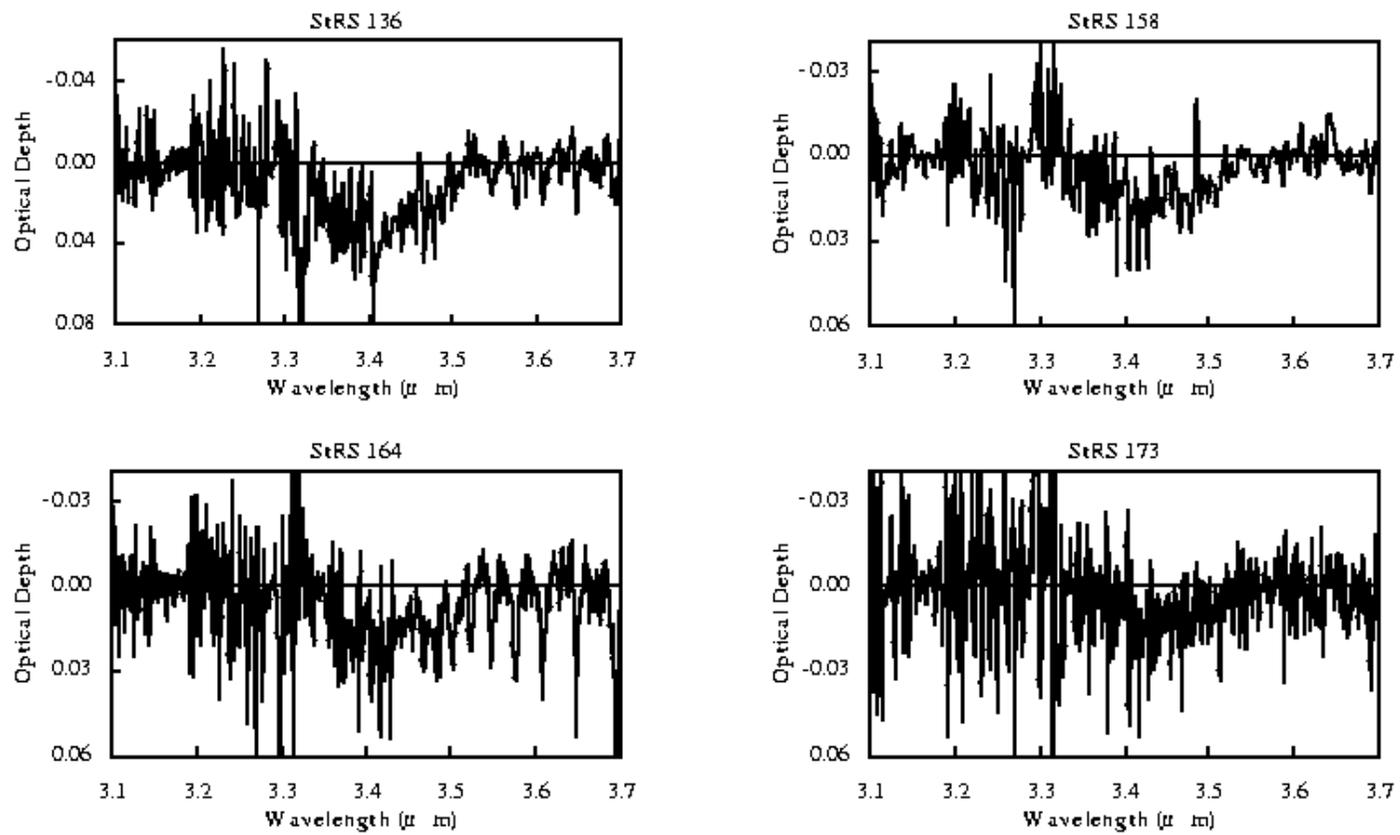

**Figure 4.** Optical depth spectra of the 3.4-$\mu$m feature towards selected genuinely early-type Stephenson stars.

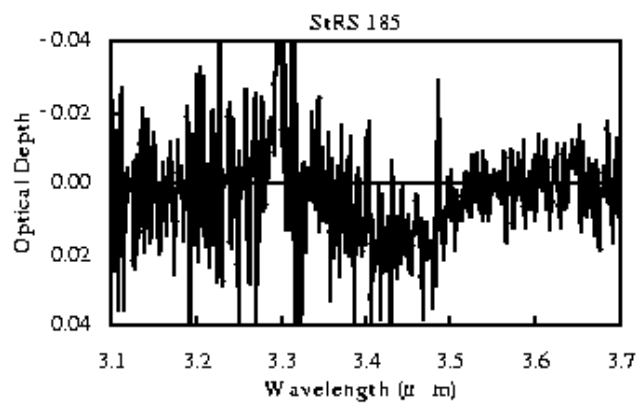
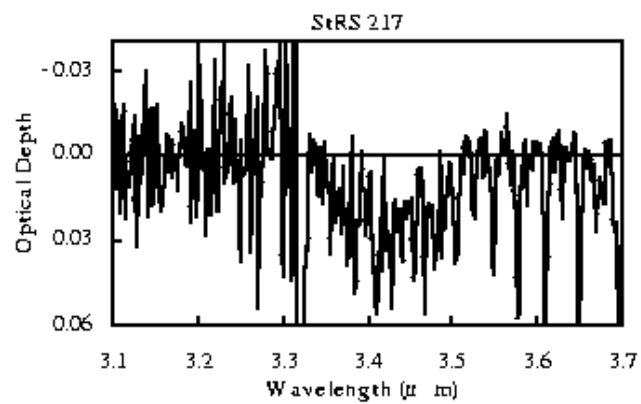
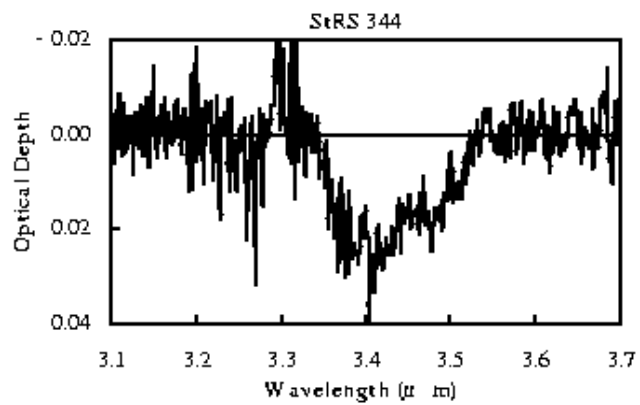
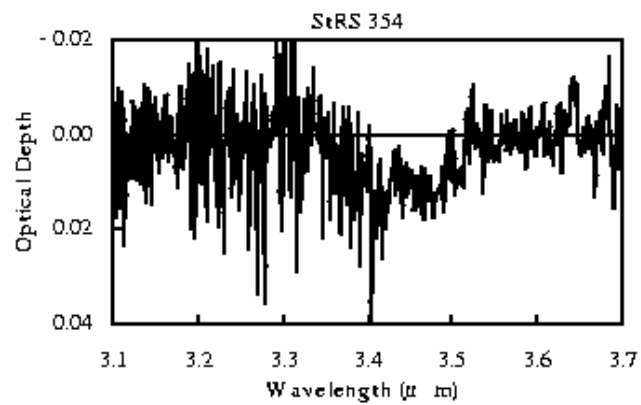

**Figure 4.** Continued

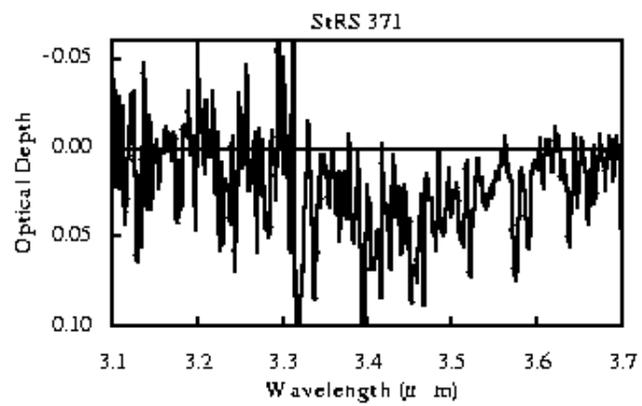
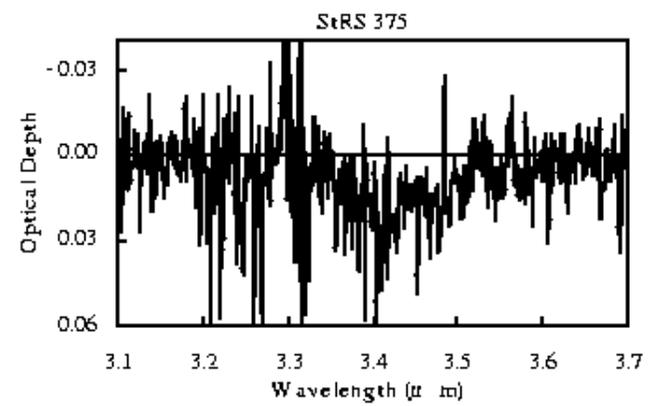
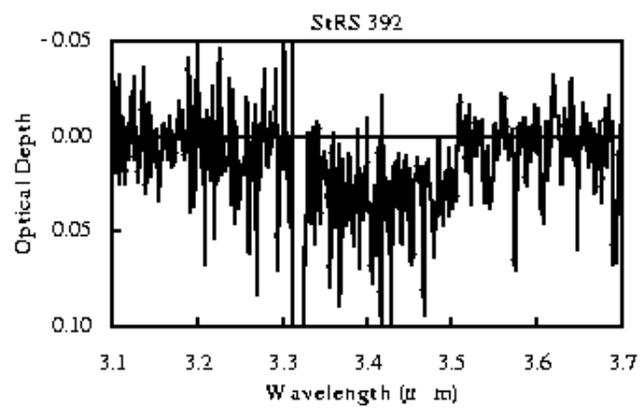
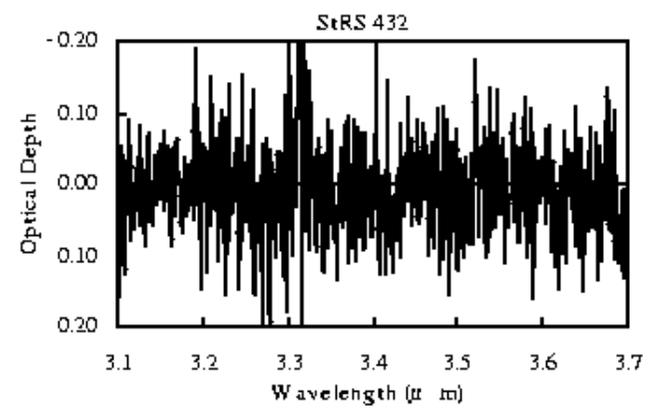

**Figure 4.** Continued

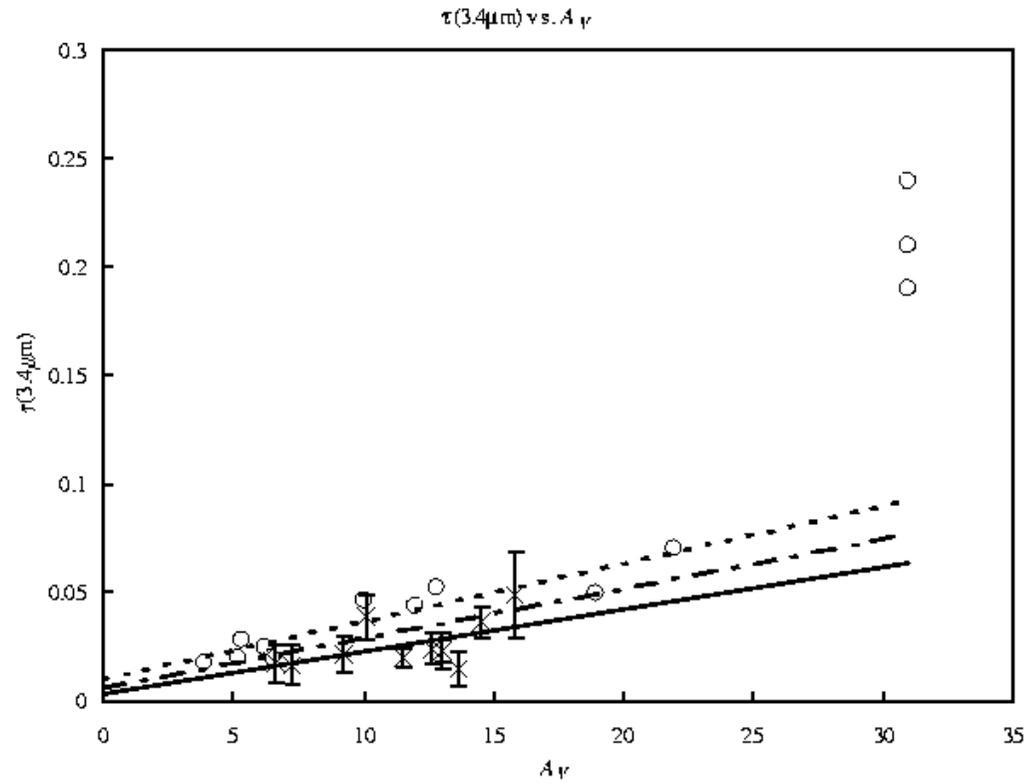

**Figure 5.** The modified $\tau$ - $A(V)$ relation, incorporating measurements of the early-type Stephenson stars (crosses) and existing data from the literature (open circles, taken from Pendleton et al. 1994). The solid line is a linear fit to the StRS objects alone ($\tau = 0.0019\ A(V) + 0.0032$), the regular dashed line is a linear fit to the Pendleton data alone ($\tau = 0.0027\ A(V) + 0.0098$), and the irregular dashed line is a fit to the combined datasets ($\tau = 0.0023\ A(V) + 0.0060$). Feature optical depth and $A(V)$ values for the StRS objects are as listed in Table 1.

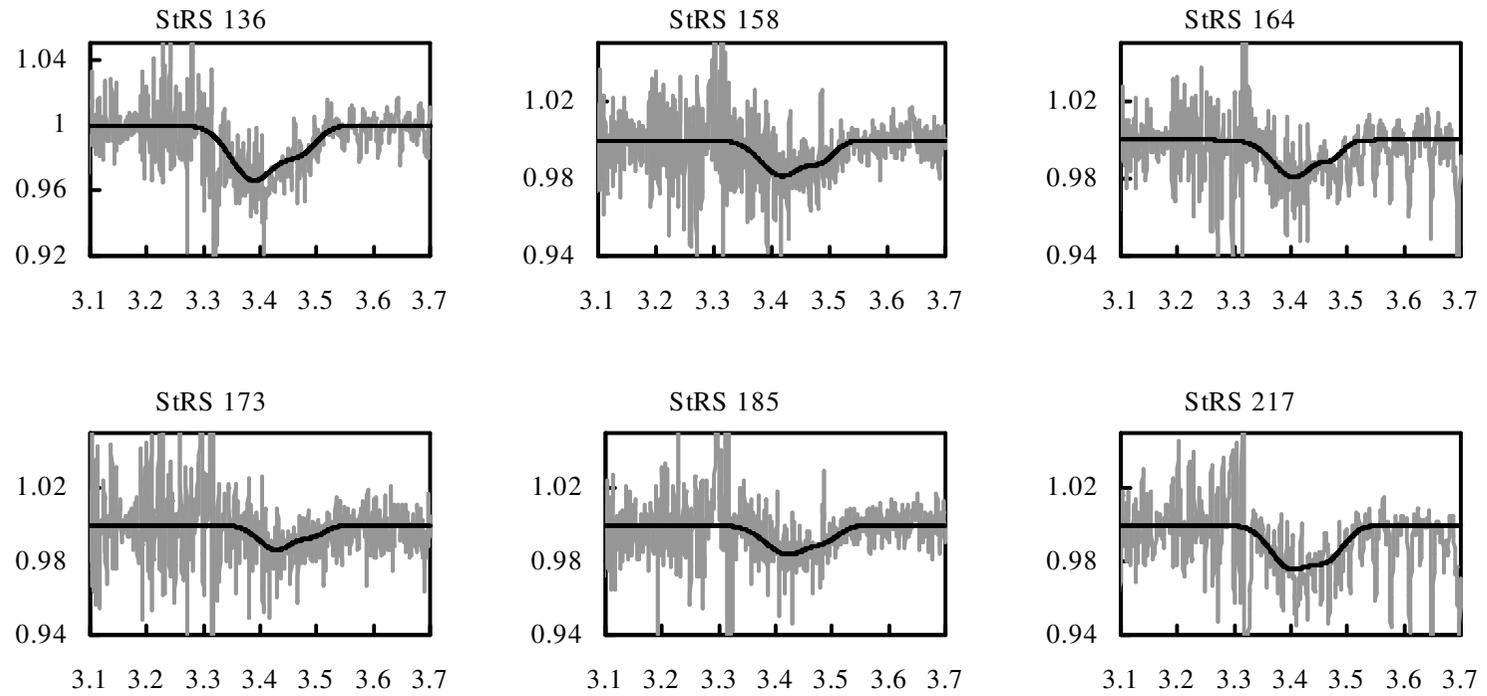

**Figure 6.** Gaussian fits to the 3.4-$\mu$m absorption band in the normalised flux spectra towards the early-type Stephenson stars. In all cases, the abscissa is the wavelength ($\mu$m) and the ordinate is the normalised flux.

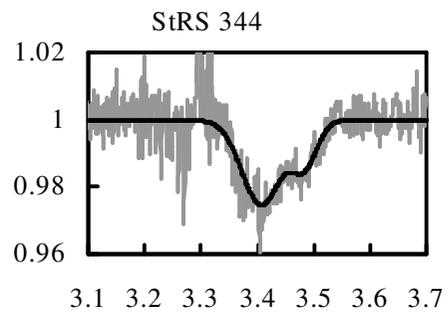 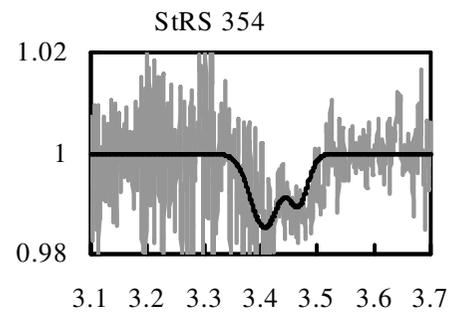 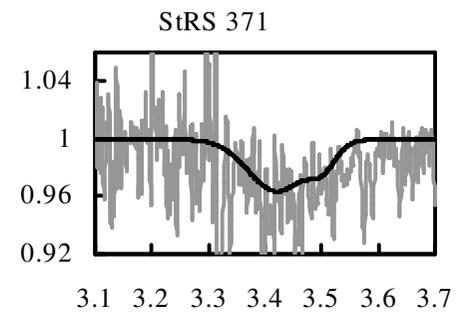
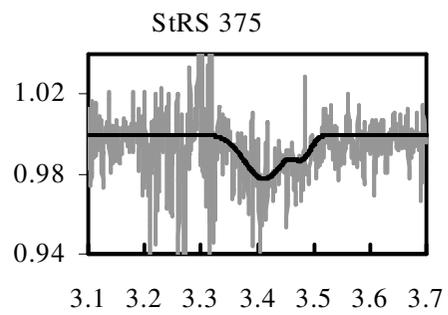 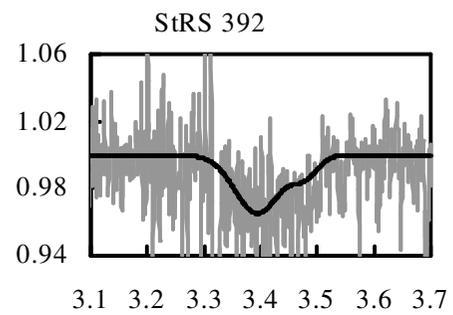

**Figure 6.** Continued

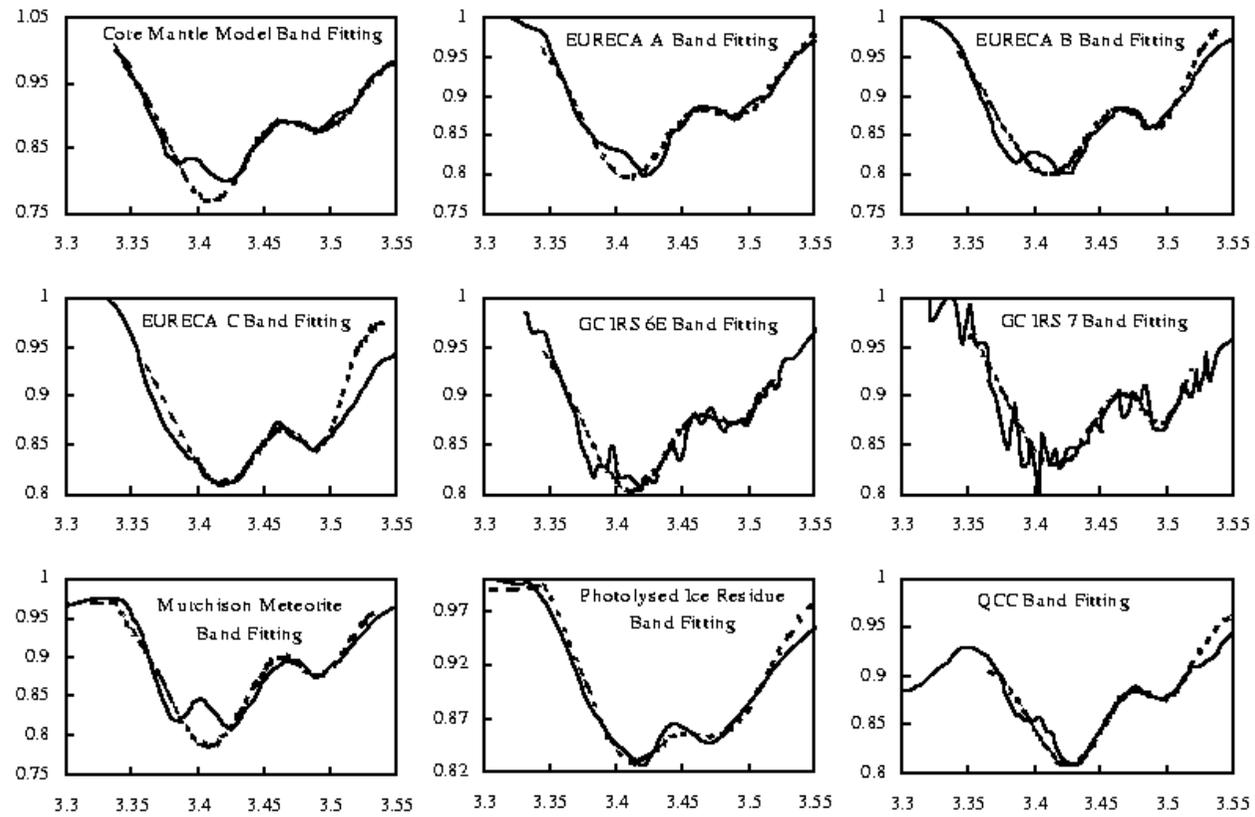

**Figure 7.** Gaussian fits (dashed lines) to the 3.4-$\mu$m absorption feature for a range of profiles obtained from the literature (solid lines). The ordinates are the wavelength ($\mu$m) and the abscissas are the normalised flux.

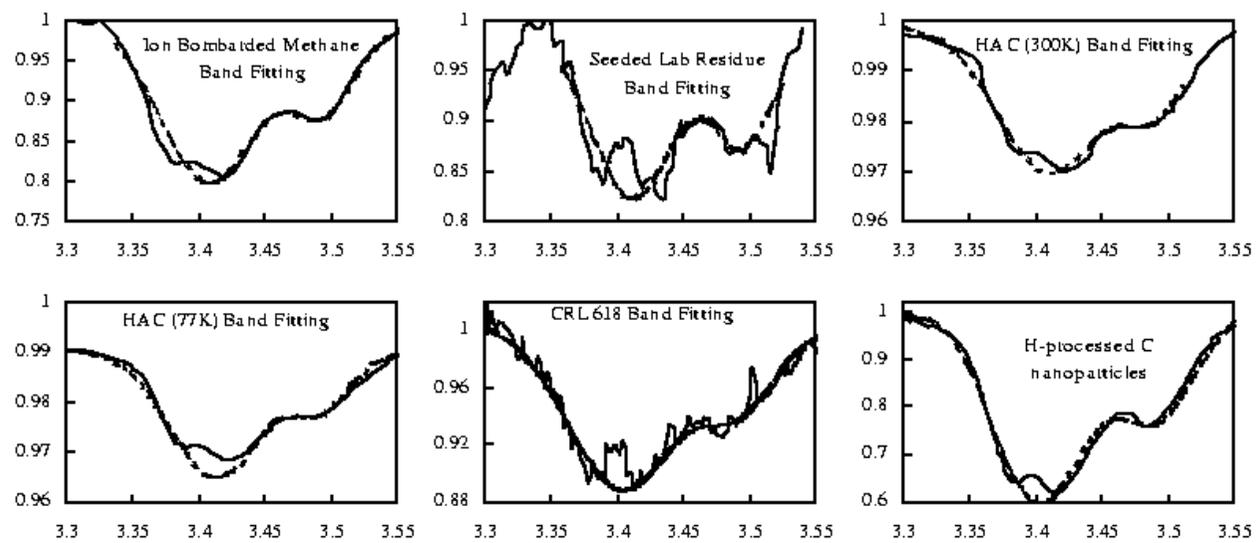

**Figure 7.** Continued

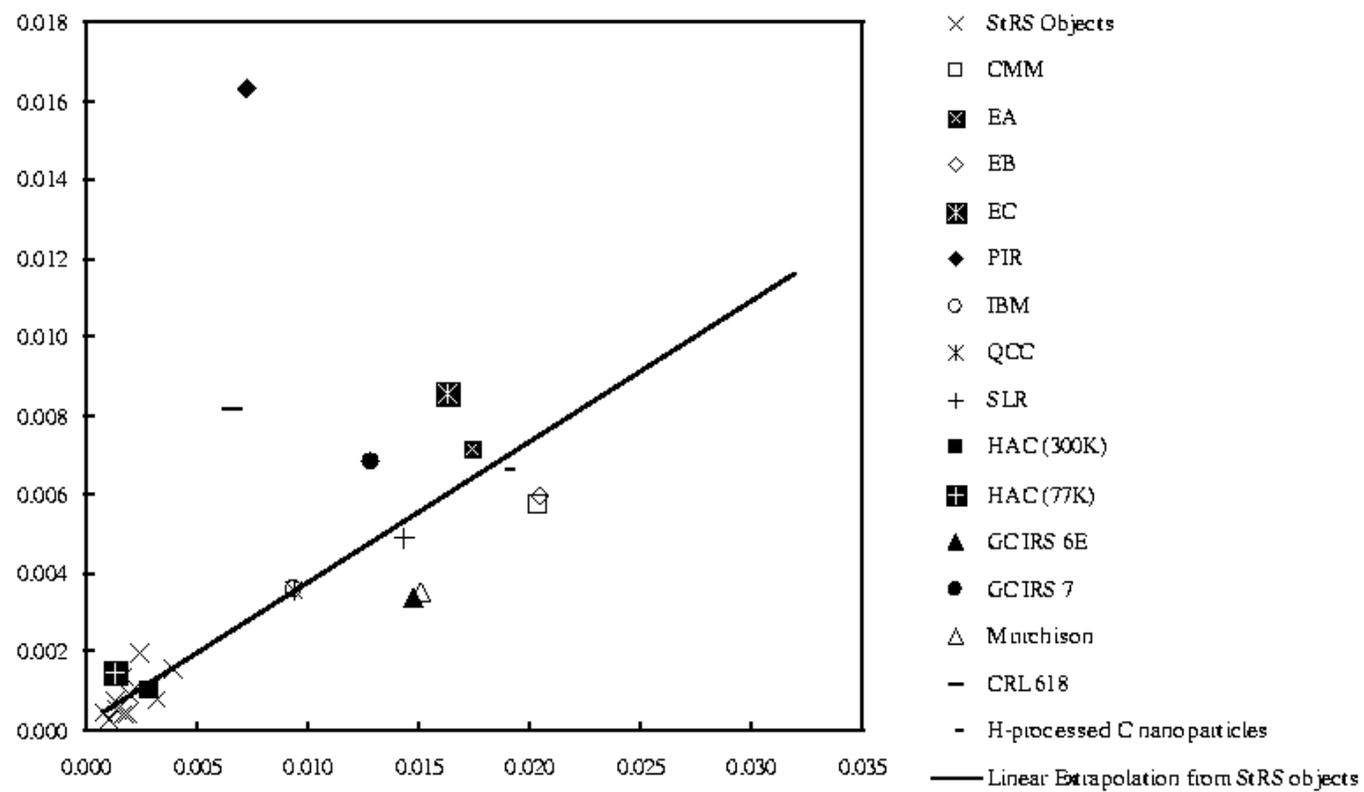

**Figure 8.** A comparison of the relative contributions of the two Gaussians used in fits to a range of 3.4-$\mu$m feature profiles.

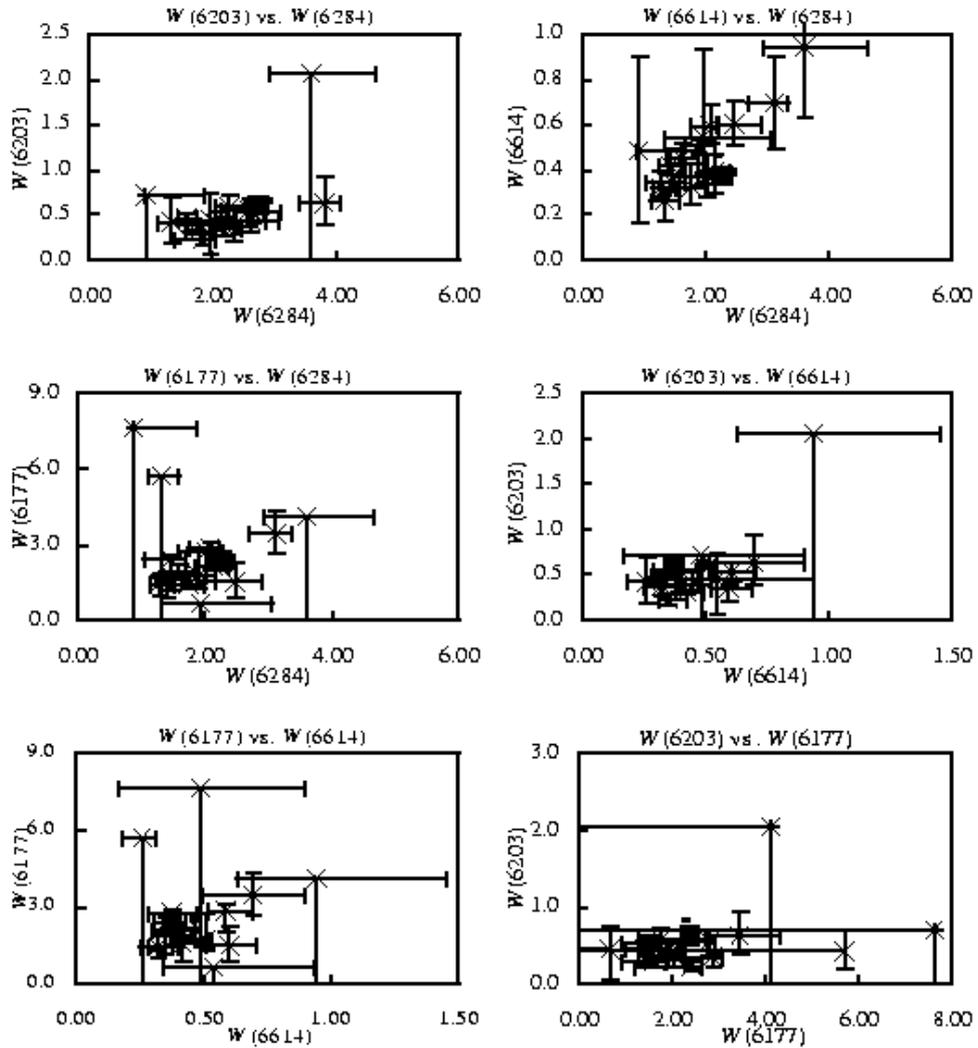

**Figure 9.** A comparison of the strength of the DIBs in the early-type Stephenson stars when not normalised by *A(V)*. Upper limiting values are shown as conventional data points, but with full-range error bars. Cyg OB 2 No. 12 is shown as a black square with dashed error bars.

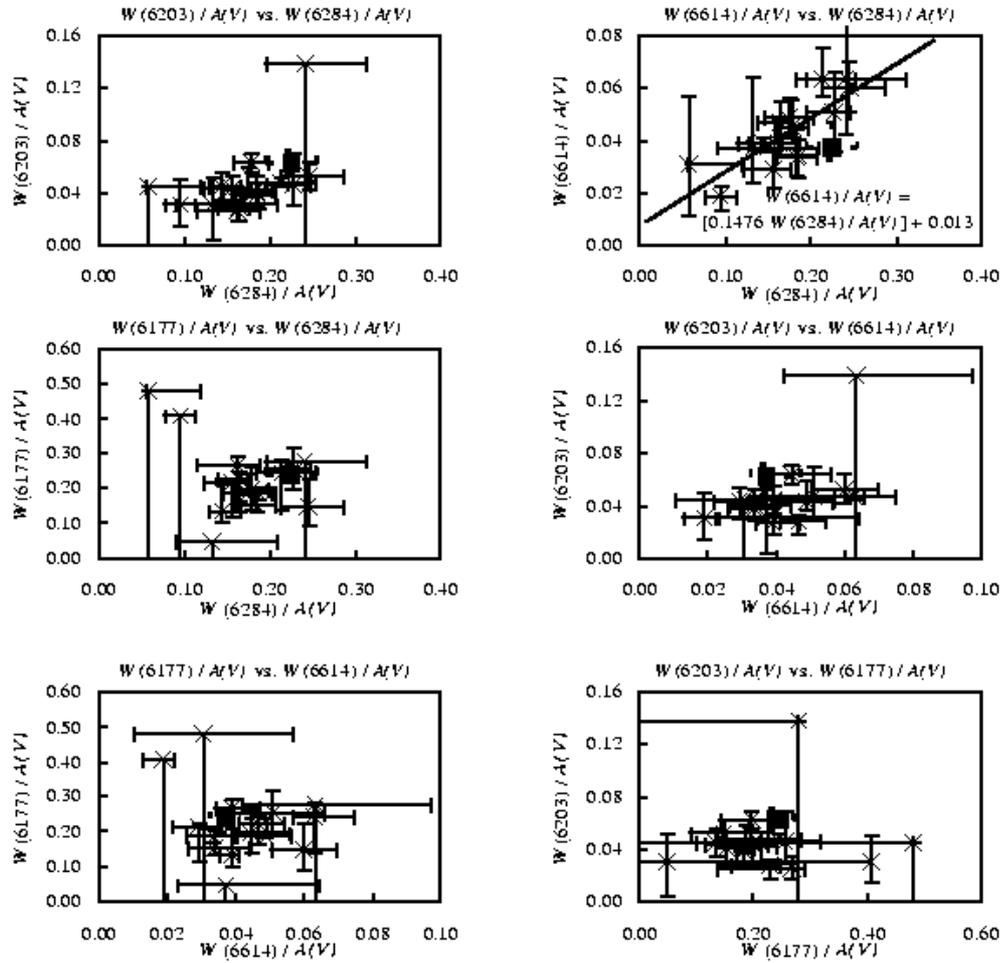

**Figure 10.** A comparison of the strength of the DIBs in the early-type Stephenson stars (crosses) when normalised by *A(V)*. Upper limiting values are shown as conventional data points, but with full-range error bars. Cyg OB 2 No. 12 is shown as a black square with dashed error bars. The solid line (with its corresponding equation) in the upper right panel is a first-order trend line fitted to the StRS data, indicating a possible DIB correlation.

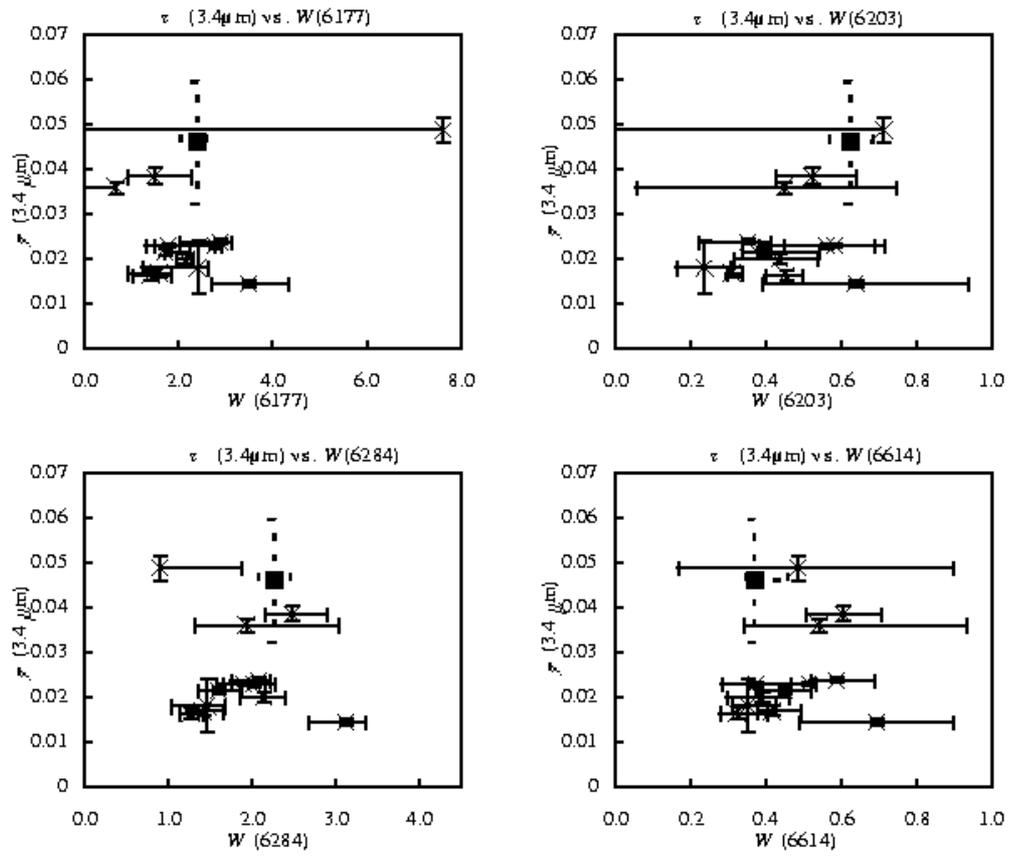

**Figure 11.** A comparison of the optical depth of the 3.4-$\mu$m C - H stretch band with the measured DIBs equivalent widths, not normalised with respect to visual extinction. The crosses are the StRS objects, and the solid square with dashed error bars is Cyg OB2 No. 12.

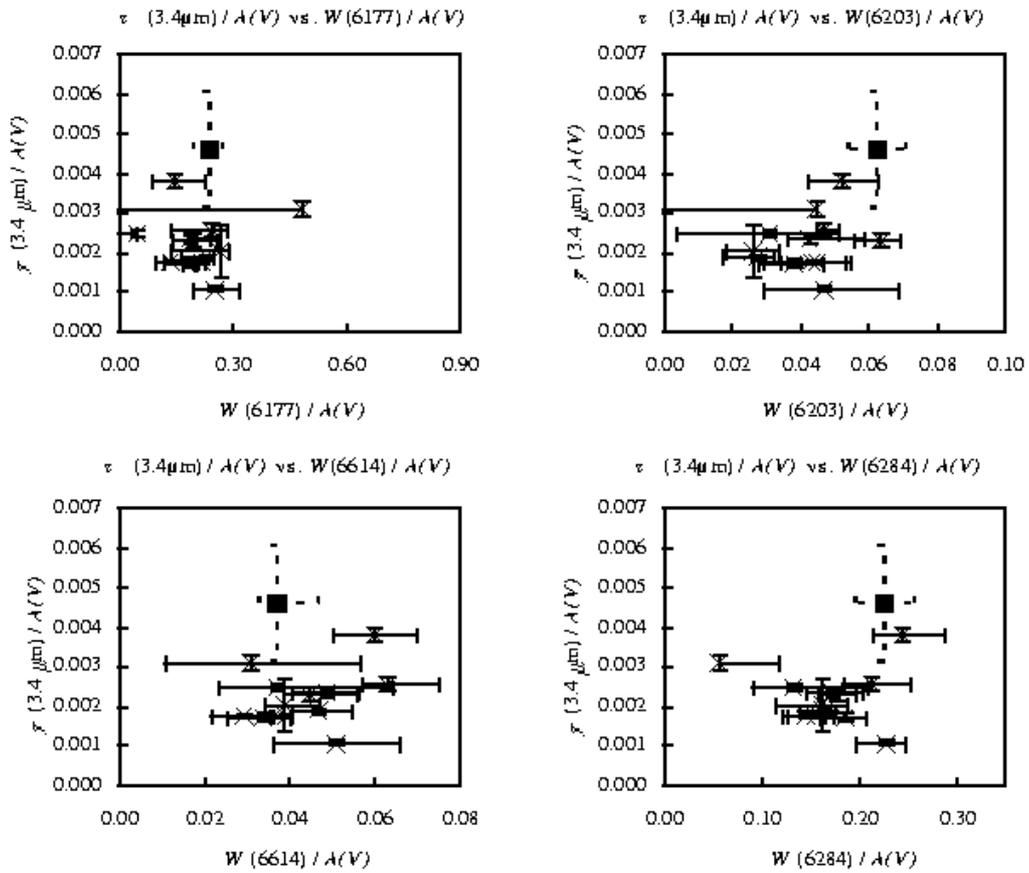

**Figure 12.** A comparison of the optical depth of the 3.4-$\mu$m C - H stretch feature with the measured DIBs equivalent widths, all normalised with respect to visual extinction. The crosses are the StRS objects, and the solid square with dashed error bars is Cyg OB2 No. 12.

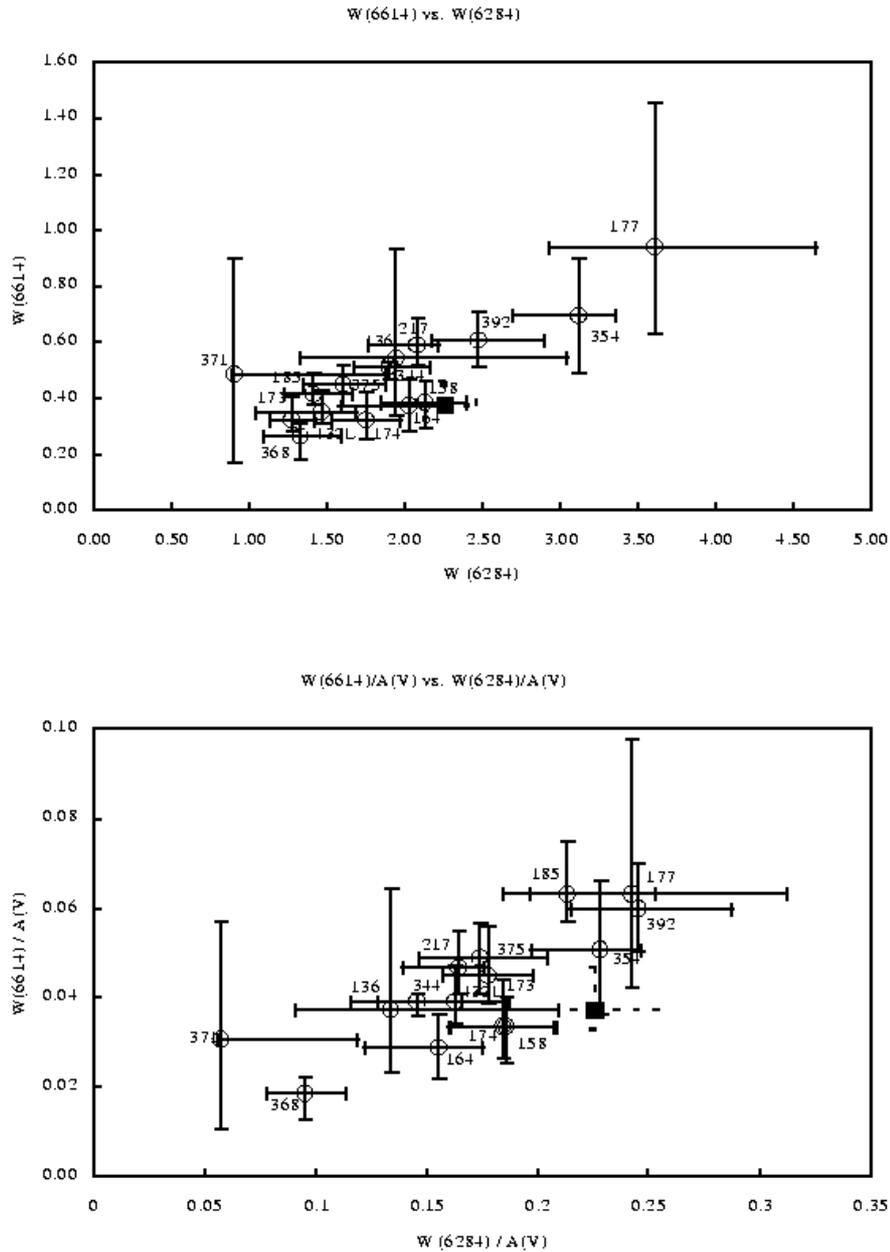

**Figure 13.** Correlation plots of λ6614 with λ6284. The upper plot shows the measured equivalent widths of the DIBs and the lower plot shows equivalent widths when normalised with respect to $A(V)$. Circles denote the StRS objects with both DIB and 3.4$\mu$m measurements, labelled with the corresponding StRS numbers. An 'L' after the StRS number denotes an upper limit only for $\tau(3.4\mu m)$. Filled and open circles in the upper plot denote stars with $\tau(3.4\mu m)$ greater and less than the sample average $\tau(3.4\mu m) = 0.0022$ respectively. Filled and open circles in the lower plot denote stars with $\tau(3.4\mu m) / A(V)$ greater and less than the sample average $\tau(3.4\mu m) / A(V) = 0.026$ respectively. Open squares denote sources lacking 3.4$\mu$m measurements. The solid square with dashed error bars is Cyg OB2 No. 12.